\documentclass[%
reprint,
showpacs,preprintnumbers,
 amsmath,amssymb,
 aps,
]{revtex4-1}
\usepackage{xcolor}
\usepackage{siunitx}
\usepackage{graphicx}
\usepackage{dcolumn}
\usepackage{bm}

\usepackage{slashed}

\newcommand{\be}{\begin{equation}}
\newcommand{\ee}{\end{equation}}

\newcommand{\muhat}{{\hat{\mu}}}

\newcommand{\CE}{{\cal E}}

\newcommand{\eps}{\epsilon}

\newcommand*\xbar[1]{%
   \hbox{%
     \vbox{%
       \hrule height 0.5pt 
       \kern0.5ex
       \hbox{%
         \kern-0.1em
         \ensuremath{#1}%
         \kern-0.1em
       }%
     }%
   }%
}

\begin{document}
\setlength{\parskip}{0pt}

\preprint{INT-PUB-19-046}

\title{Pions in hot dense matter and their astrophysical implications}

\author{Bryce Fore and Sanjay Reddy}
\affiliation{%
Department of Physics, University of Washington, Seattle, WA 98195
}%

\affiliation{%
Institute for Nuclear Theory, University of Washington, Seattle, WA 98195}%

\date{\today}

\begin{abstract}
We study the role of pions in hot dense matter encountered in astrophysics. We find that strong interactions enhance the number density of negatively charged pions and that this enhancement can be calculated reliably for a relevant range of density and temperature using the virial expansion. We assess the influence of pions and muons on the equation of state (EOS) and weak interaction rates in hot dense matter. We find that thermal pions increase the proton fraction and soften the EOS. We also find that charged current weak reactions involving pions and muons $\xbar{\nu}_\mu+\mu^- \rightarrow \pi$ and $\nu_\mu+\pi^- \rightarrow \mu^-$ make an important contribution to the opacity of muon neutrinos. This could influence the dynamics of core-collapse supernovae and neutron star mergers. Finally, we note that pion-nucleon reactions can alter the evolution of the proton fraction when weak interactions are not in equilibrium. 
\end{abstract}

\maketitle


\section{\label{sec:intro}Introduction}

Hot dense matter encountered inside neutron stars plays an essential role in the dynamics of extreme astrophysical phenomena such as core-collapse supernovae and neutron star mergers. Properties of nuclear matter, especially the equation of state (EOS), at the high density ($10^{11}-10^{15}$ g/cm$^3$) and temperature (T=5-50 MeV) realized in these astrophysical sites have been studied in some detail (for a recent review see \cite{Oertel:2016bki}). However, the role of pions in core-collapse supernovae and neutron star mergers is not well understood. Motivated by recent work in Ref.~\cite{muon_nova} which showed the inclusion of muons could facilitate supernova explosions through the neutrino-driven mechanism, we study the role of negatively charged thermal pions and muons in the hot neutron-rich matter and discuss their implications. 

The neutron-rich dense stellar matter, which is electrically neutral and typically in beta-equilibrium, contains a high density of electrons. The electron chemical potential, $\mu_e$, increases with density and acts as a source for other negatively charged particles. At low temperature, muons appear when $\mu_e > m_\mu$ where $m_\mu=105.7$ MeV is the muon mass. In the absence of interactions between pions and nucleons, a Bose-Einstein condensate of negatively charged pions would appear when $\mu_e > m_{\pi^-}$ where $m_{\pi^-}=139.6$ MeV is the mass of the charged pion. Since the electromagnetic interactions of muons in the dense matter are negligible, interactions do not alter the threshold density for their appearance. In contrast, pions interact strongly with nucleons. Their dispersion relation in a dense medium is poorly understood as it is sensitive to nuclear many-body effects that are difficult to calculate reliably at high density. In the early 1970s, Migdal and Sawyer independently proposed that pions could condense in dense matter due to attractive p-wave interactions with nucleons \cite{Migdal:1973, Sawyer:1972}. Several other studies have explored in some detail the possibility of pion condensation in the dense neutron-rich matter (see for example Refs. \cite{Migdal:1978, Weise:1975, Backman:1975, Migdal:1990, Akmal:1997ft, Akmal:1997ft}). Despite these studies, the type of condensation and the critical density for its appearance remain uncertain. In this article, we restrict our attention to high temperatures and low density where we expect a population of thermal pions. Under these conditions, the virial expansion provides a reliable approach to include interactions between pions and nucleons.  We find that these interactions enhance the thermal population of pions and influence the composition of matter, the EOS, and weak interaction rates.
  
We begin by discussing a simple model that describes hot dense nuclear matter containing pions and muons in section \ref{sec:eos}. Here, the virial expansion is used to include the effects of pion-nucleon interactions, and nucleon-nucleon interactions are accounted for by a phenomenological mean-field model. In section \ref{sec:implications}, we discuss the effect of pions and muons on the equation of state, the neutrino opacity, and on the processes relevant to understanding bulk viscosity of dense matter. Finally, in section \ref{sec:conclusion}, we conclude and identify future work needed to properly include pions in the description of core-collapse supernovae and neutron star mergers.  

\section{\label{sec:eos}Pions in hot neutron-rich matter}
 At the temperatures of interest, in the range $10-50$ MeV, the matter is composed of nucleons, leptons and pions. To include the effects of interactions between pions and nucleons, we calculate the second-virial coefficient for the pion-nucleon system directly in terms of the measured pion-nucleon phase shifts. This approach was used to describe the hot hadronic gas encountered in heavy-ion collisions in Ref.~\cite{Dashen:1969, Venugopalan:1992hy, Huovinen:2017ogf}, and to describe a dilute gas of nucleons encountered in outer regions of the newly born neutron star in Ref.~\cite{Horowitz:2005zv}. The virial expansion provides a systematic approach to calculate the thermodynamic properties of interacting multi-component gases when the particle fugacities are small. The fugacity of a particle species $i$ is given by $z_i=\exp{\beta(\mu_i-m_i)}$ where $\mu_i$ is the chemical potential which includes mass, $m_i$ is the rest mass of the particle, $\beta=1/T$ is the inverse temperature. To justify the use of the virial expansion, we restrict our analysis to densities that are low enough, and temperatures that are high enough, to ensure that the fugacity $z_{\pi^-} < 1$ and that Bose-Einstein condensation of pions does not occur.

At the modest densities that we consider, $\rho \lesssim 3 \times 10^{14}$ g/cm$^3$, it is adequate, as a first step, to account for interactions between nucleons using a simple non-relativistic Skyrme model \cite{Skyrme:1958}. The parameters of the model we employ, called NRAPR, are obtained by fitting to the empirical properties of nuclear matter at nuclear saturation density $n_0=0.16$ fm$^{-3}$ \cite{Steiner:2004fi, Dutra:2012mb}, and to the properties of neutron matter predicted by ab initio many-body theory which employ realistic nuclear interactions \cite{Tews:2012fj,Gandolfi:2013baa}. The nucleon contribution to the energy density is given by 
 \begin{equation}
 \begin{split}
 \CE_N & (n_n,n_p,T)= \frac{\tau_n}{2 m_n} + \frac{\tau_p}{2 m_p} \\
 & + n_B(\tau_n+\tau_p) \left[\frac{t_1}{4}\left(1+\frac{x_1}{2}\right)+\frac{t_2}{4}\left(1+\frac{x_2}{2}\right)\right]
 \\& +(\tau_n n_n+\tau_p n_p) \left[\frac{t_2}{4}\left(\frac{1}{2}+x_2\right)-\frac{t_1}{4}\left(\frac{1}{2}+x_1\right)\right]
 \\& +\frac{t_0}{2}\left[\left(1+\frac{x_0}{2}\right)n_B^2-\left(\frac{1}{2}+x_0\right)(n_n^2+n_p^2)\right]
 \\& +\frac{t_3}{12}\left[\left(1+\frac{x_3}{2}\right)n_B^2-\left(\frac{1}{2}+x_3\right)(n_n^2+n_p^2)\right]n^\eps]\,, 
 \end{split}
 \label{eq:enucleon} 
 \end{equation}
where $t_0, t_1, t_2, t_3, x_0, x_1, x_2, x_3,$ and $\eps$ are the Skryme parameters taken from Ref.~\cite{Steiner:2004fi}. The neutron and proton densities are denoted by $n_n$ and $n_p$, respectively, and $n_B=n_n+n_p$ is the total baryon density. The variables $\tau_n$ and $\tau_p$ are defined such that the first two terms in Eq.~\ref{eq:enucleon} correspond to the neutron and proton kinetic energy densities, respectively.

The dense matter we consider is homogeneous, electrically neutral, and close to beta equilibrium. Under these conditions, the chemical potential for negative charge $\hat{\mu}= \mu_n-\mu_p$ acts as a source for negatively charged particles. In beta-equilibrium the electron, muon and pion chemical potentials are equal $\mu_e=\mu_\mu^- = \mu_{\pi^-}= \hat{\mu}$, and electric charge-neutrality requires that $n_p = n_e+n_\mu+n_{\pi^-}$. When $\hat{\mu}=\mu_n-\mu_p \gtrsim T$ it is reasonable to neglect the presence of $\pi^0$ and $\pi^+$ particles in the ground state since their density is suppressed by the factor $\exp{(-\hat{\mu}/T)}$ and $\exp{(-2\hat{\mu}/T)}$, respectively, relative to the abundance of $\pi^-$.  

The second-virial coefficient for the $\pi^- -$neutron system is given by 
\begin{equation} 
b^{n\pi^-}_{2}=\frac{e^{\beta M}}{2\pi^3} \int_M^\infty dE E^2 K_1(\beta E) \sum_{l,\nu} (2l+1) \delta^{3/2}_{l,\nu}\,,
\label{eq:virialnpi} 
\end{equation} 
where $K_1$ is the modified Bessel function of the second kind, $M=m_N+m_\pi$ is the invariant mass of the interacting pair at the threshold. This result was obtained  using the relativistic formalism in Ref.~\cite{Dashen:1969,Venugopalan:1992hy} and is appropriate for our study because pions can be relativistic as their typical pion momenta $p_\pi \simeq  \sqrt{3 m_\pi T}$ is comparable to $m_\pi$. Note that the phase shifts $\delta$ are dependent on $E$ which is the center of mass energy. The sum is over the angular momentum $l$ of the scattering state, and the nucleon spin-projections $\nu=+,-$. Since $n\pi^-$ scattering only involves the isospin $I=3/2$ state, only the pion-nucleon phase shift in the isospin $I=3/2$ channel denoted by $\delta^{3/2}_{l,\nu} $ contributes to $b^{n\pi^-}_{2}$. We note that this definition differs from Ref.~\cite{Dashen:1969,Venugopalan:1992hy} in that it contains an extra factor of $e^{\beta M}$. We find it convenient to include this factor and redefine the thermodynamic functions that appear later in the text.

Proton-$\pi^-$ scattering involves two reaction channels: $\pi^- p \rightarrow \pi^- p$ and $\pi^-p \rightarrow \pi^0 n$, which implies these reactions do not have definite isospin. However, since $b_2$ is independent of the basis in which we consider the S matrix \cite{Lo:2017}, $b_2^{p\pi^-}$ depends on the sum of the phase shifts from the two mixed isospin channels.

\begin{equation}  
b^{p\pi^-}_{2}=\frac{e^{\beta M}}{2\pi^3} \int_M^\infty dE E^2 K_1(\beta E) \sum_{l,\nu} (2l+1) (\delta^{3/2}_{l,\nu} + \delta^{1/2}_{l,\nu} ) \,. 
\label{eq:virialppi} 
\end{equation}

\begin{figure}[h]
\includegraphics[width=.45\textwidth]{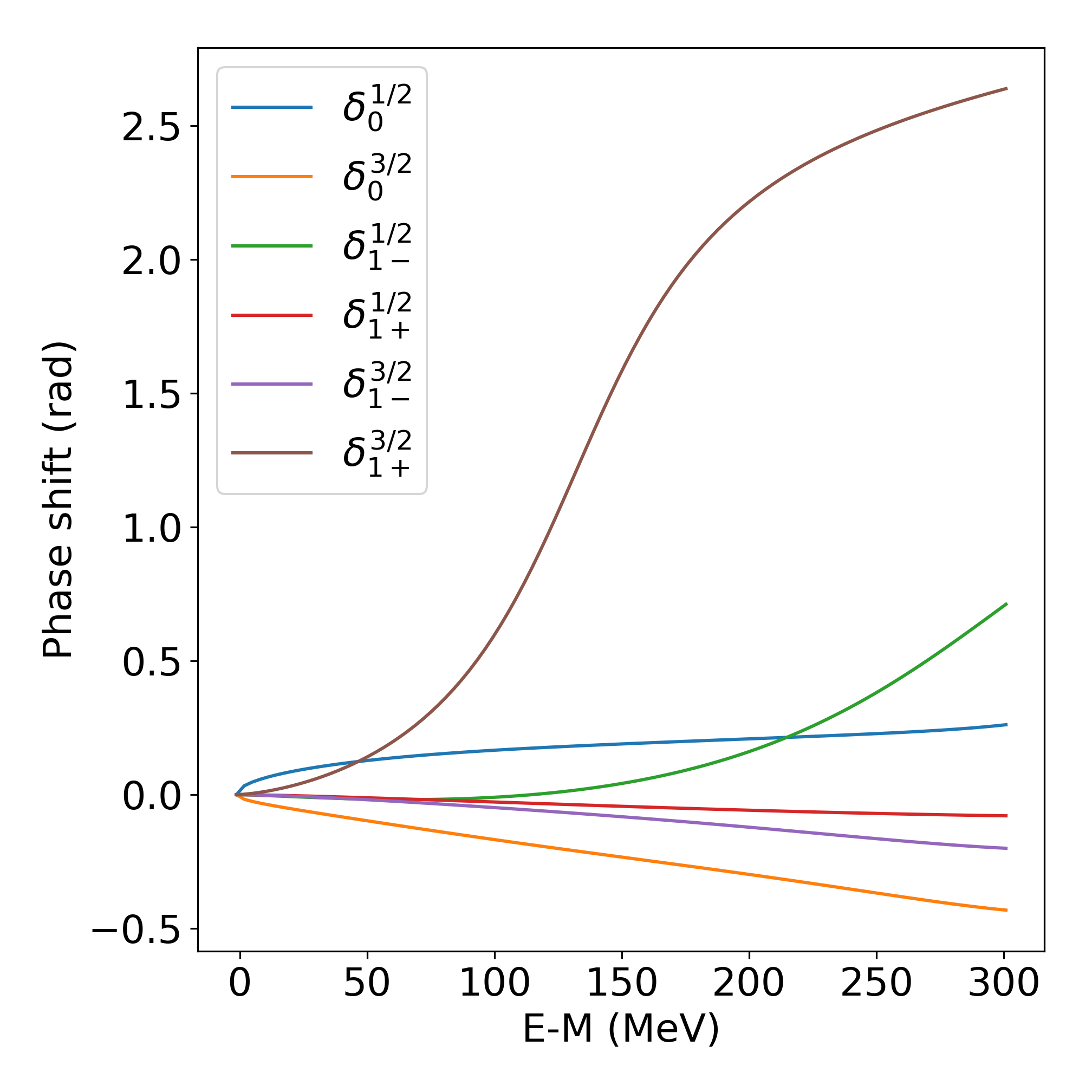}
\caption{Pion-nucleon phase shifts from the analysis presented in Ref.~\cite{Hoferichter:2015hva}}
\label{fig:phase_shifts}
\end{figure}

In this study we will only include $l=0,1$, i.e, the $s$-wave and $p$-wave contributions. At the energies of interest we find them to be the dominant contributions. In Fig.~\ref{fig:phase_shifts} we show the $s$-wave and $p$-wave phase shifts taken from the analysis of experimental data in Ref.~\cite{Hoferichter:2015hva}. The phase shifts are plotted as a function of $E-M$ where $E= \sqrt{p^2+m^2_\pi} + \sqrt{p^2+m_N^2}$ is the center of mass energy and $p$ is the magnitude of the pion and nucleon momenta in the center of mass frame. The large and attractive p-wave phase-shift $\delta^{3/2}_{+1}$ due to the $\Delta-$resonance is the dominant channel. 
The second-virial coefficients calculated using Eqns. \ref{eq:virialnpi} \& \ref{eq:virialppi} at a few temperatures of interest are shown in Table 1.
\begin{table}[]
\begin{tabular}{|l|c|c|c|}
\hline
 T (MeV) & 15 & 30 & 45 \\ \hline
 $b_2^{n\pi^-}$ (fm$^{-3}$)& \num{2.14E-4} & \num{4.09E-3} & \num{1.87E-2} \\ \hline
 $b_2^{p\pi^-}$ (fm$^{-3}$)& \num{4.24E-4} & \num{4.68E-3} & \num{2.02E-2} \\ \hline
\end{tabular}
\caption{The second-virial coefficients for the $N\pi^-$ system.}
\end{table}

The virial expansion for nucleons fails at the higher density of interest here, and for this reason we use a simple mean-field model to include the effects of nucleon-nucleon interactions. While it is desirable to treat the nucleon-nucleon and nucleon-pion interactions consistently, and chiral perturbation theory provides in-principle a framework to do this, there remain technical challenges \cite{Meissner:2001gz}. Further, the convergence of the chiral expansion for pion-nucleon interactions is poor and requires a large number of operators to capture the resonant nature of this interaction\cite{Hoferichter:2015tha}. To circumvent these issues, as a first step in the study of the role of pions in hot dense matter, we advocate our hybrid approach. In the limit when $z_{\pi^-} \ll 1$, and $z_n,z_p \ll1$ our approach is reliable. At higher density where the $z_{\pi^-} < 1/2$, and $z_n < 1$ or $z_p < 1$ we expect our results to capture the qualitative aspects, but corrections due to neglected terms proportional to $z_\pi z^2_n$ and $z_\pi z^2_p$ become important. These need to be assessed before one can draw quantitative conclusions. In this study, we also neglect pion-pion interactions because the pion-nucleon interaction, and the nucleon density, are both significantly larger. 

The composition of matter at fixed temperature and baryon density is determined by requiring matter to be charge-neutral and in beta-equilibrium, The chemical potentials $\mu_n$, $\mu_p$, and $\mu_e=\mu_{\mu^-}=\mu_{\pi^-} = \muhat = \mu_n-\mu_p$ are determined to ensure that $n_B=n_n+n_p$ and $n_{e^-} + n_{\mu^-} + n_{\pi^-}=n_p$. The effect of interactions is negligible for the leptons and their number densities are obtained using the ideal Fermi gas result. 

For nucleons and pions, interactions are important. The nucleon number densities are given by 
\begin{equation} 
n_i=\int \frac{dk}{\pi^2} k^2 (1+\exp{(\beta(\epsilon_i(k)-\mu_i)}))^{-1} \,. 
\label{eq:nucleon_density} 
\end{equation}   
where the single nucleon energy 
\begin{equation} 
\epsilon_i(k)= m_i + \frac{k^2}{2 m^*_i} + U_i(n_n,n_p,T) \,, 
\label{eq:nucleon_energy} 
\end{equation}
is obtained in mean field theory and $m_i^*$ is the nucleon effective mass, and $U_i(n_n,n_p,T) = \partial \CE_N (n_n,n_p,T)/\partial n_i$ is the mean field potential energy \cite{Skyrme:1958}. The effective mass is found by solving for the momentum-dependent part of the functional derivative of $\CE_N (n_n,n_p,T)$ with respect to the nucleon distribution function. The momentum-independent part is equal to $U_i(n_n,n_p,T)$.

The number density of pions is obtained in the virial expansion, and is given by 
\begin{equation} 
n_{\pi^-} = \int \frac{dk}{2\pi^2} k^2 \exp{ (-\beta ( \sqrt{k^2+m^2_\pi}-\hat{\mu}))} + n^{\rm int}_{\pi^-} \,,
\label{eq:pion_number}
\end{equation} 
where 
\begin{equation}
n^{\rm int}_ {\pi^-} = \sum_{N=n,p} z_N z_{\pi^-} b_2^{N \pi^-}\,,  
\end{equation}
is the contribution due to pion-nucleon interactions. Note that the reason we have used the Boltzmann distribution here for pions rather than the Bose-Einstein distribution is for consistency with the virial expansion. The difference this makes, however, is minimal with only a change of about 4 to 5 percent in the pion number density at nuclear density and $T=30$ MeV. Eq.~\ref{eq:nucleon_density} only includes effects due to nucleon-nucleon interactions. The contributions due to pion-nucleon interactions, given by the virial expansion, are $\delta n_n = z_n z_{\pi^-} b_2^{n \pi^-}$, and $\delta n_p = z_p z_{\pi^-} b_2^{p \pi^-}$, respectively. In our hybrid model, these contributions are added to Eq.~\ref{eq:nucleon_density} to obtain the total neutron and proton densities.

\begin{figure}[h]
\includegraphics[width=.45\textwidth]{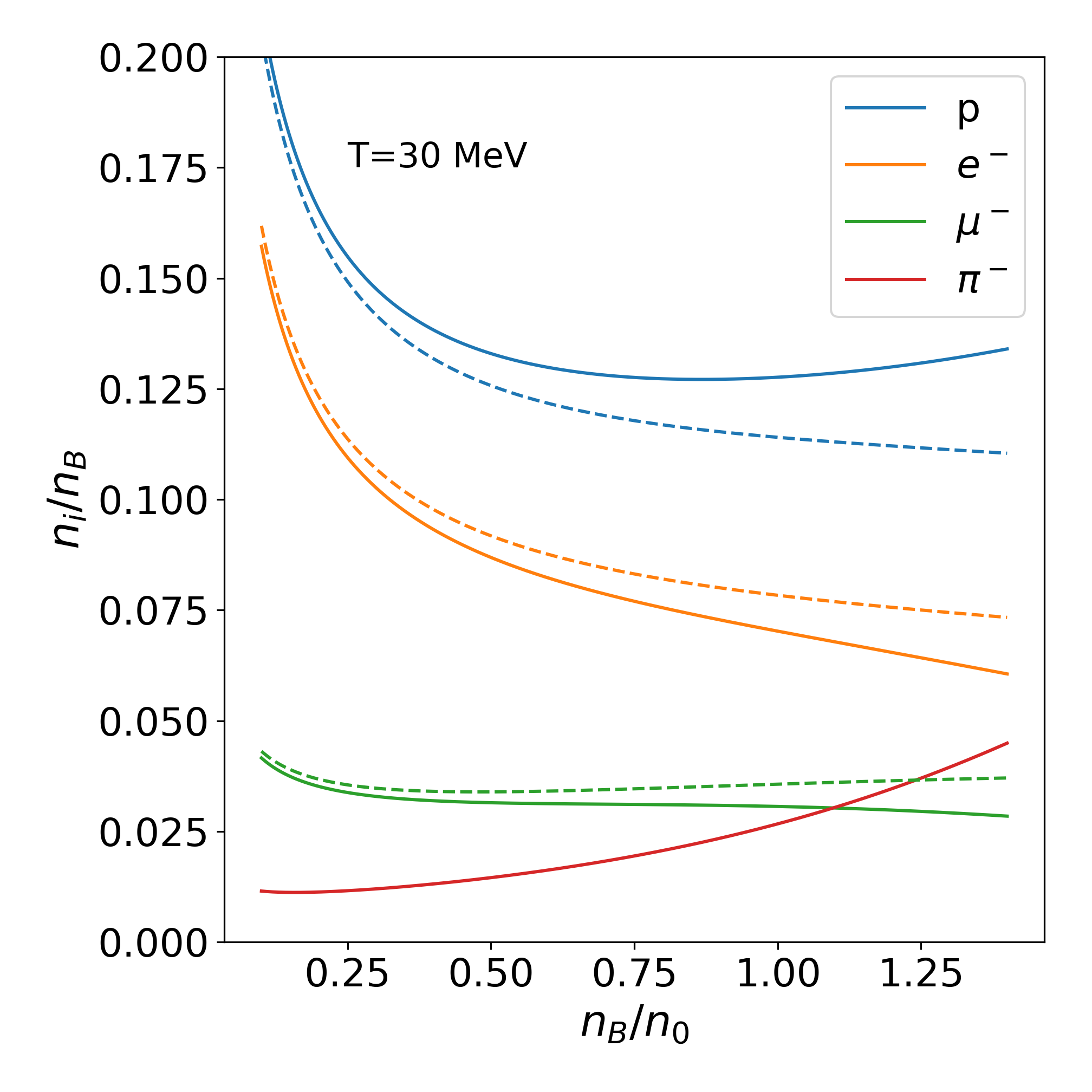}
\caption{Number fraction of charged particles at $T=30$ MeV in $\beta$-equilibrium. Solid curves include pions and dashed curves only contain nucleons and leptons.}
\label{fig:numbers}
\end{figure}

For a given value of the baryon density, $n_B$ and temperature, $T$, we guess a value for proton number density, $n_p$, and use this to define the single particle nucleon energies defined in Eq.~\ref{eq:nucleon_energy}. Then, we use  Eq.~\ref{eq:nucleon_density} to obtain the nucleon chemical potentials $\mu_n$ and $\mu_p$.  The beta-equilibrium condition allows us to obtain the charge chemical potential $\mu_e$ and we use it to obtain the number densities of $n_e, n_\mu, n_\pi$. The lepton number densities are obtained using the the ideal Fermi-Dirac distribution and the pion number density is obtained using Eq.~\ref{eq:pion_number}. The charge neutrality condition $n_p=n_e+n_\mu+n_\pi$ allows us to update the guess $n_p$ and find the true value. The above method is augmented slightly to include the change in the nucleon number densities due to the interactions between nucleons and pions. In this case we define two new variables 
\begin{equation}
\eta_n=\frac{\mu_n-m_n-U_n}{T}\,, \quad \eta_p=\frac{\mu_p-m_p-U_p}{T}\,,
\end{equation}
which we solve for in addition to the $n_p$.  These three variables are determined as a solution the system of three equations, given by  $n_n=\tilde{n}_n + \delta n_n$, $n_p=\tilde{n}_p + \delta n_p$ and $n_p=n_e+n_\mu+n_\pi$, where $\tilde{n}_n$ and $\tilde{n}_p$ are given by Eq.~\ref{eq:nucleon_density}. 

The densities of charged particles with and without the inclusion of pions are shown in Fig.~\ref{fig:numbers}. From the figure it is evident that pions enhance the proton fraction and suppress the lepton fraction in the hot dense matter as they furnish additional negative charge. This effect is strong enough that at higher densities the proton fraction begins increasing with density due to the large number of pions. Although $m_\pi > m_\mu$, strong attractive p-wave interactions with nucleons enhance their number density, at $n_B=n_0$ and $T=30$ MeV, $n_{\pi^-} \approx n_{\mu^-}$. A naive extrapolation suggests $n_{\pi^-}$ increases rapidly with density, and $n_B=1.4n_0$ and $T=30$ MeV, $n_{\pi^-} \approx 2n_{\mu^-}$. 

\begin{figure}[h]
\includegraphics[width=.45\textwidth]{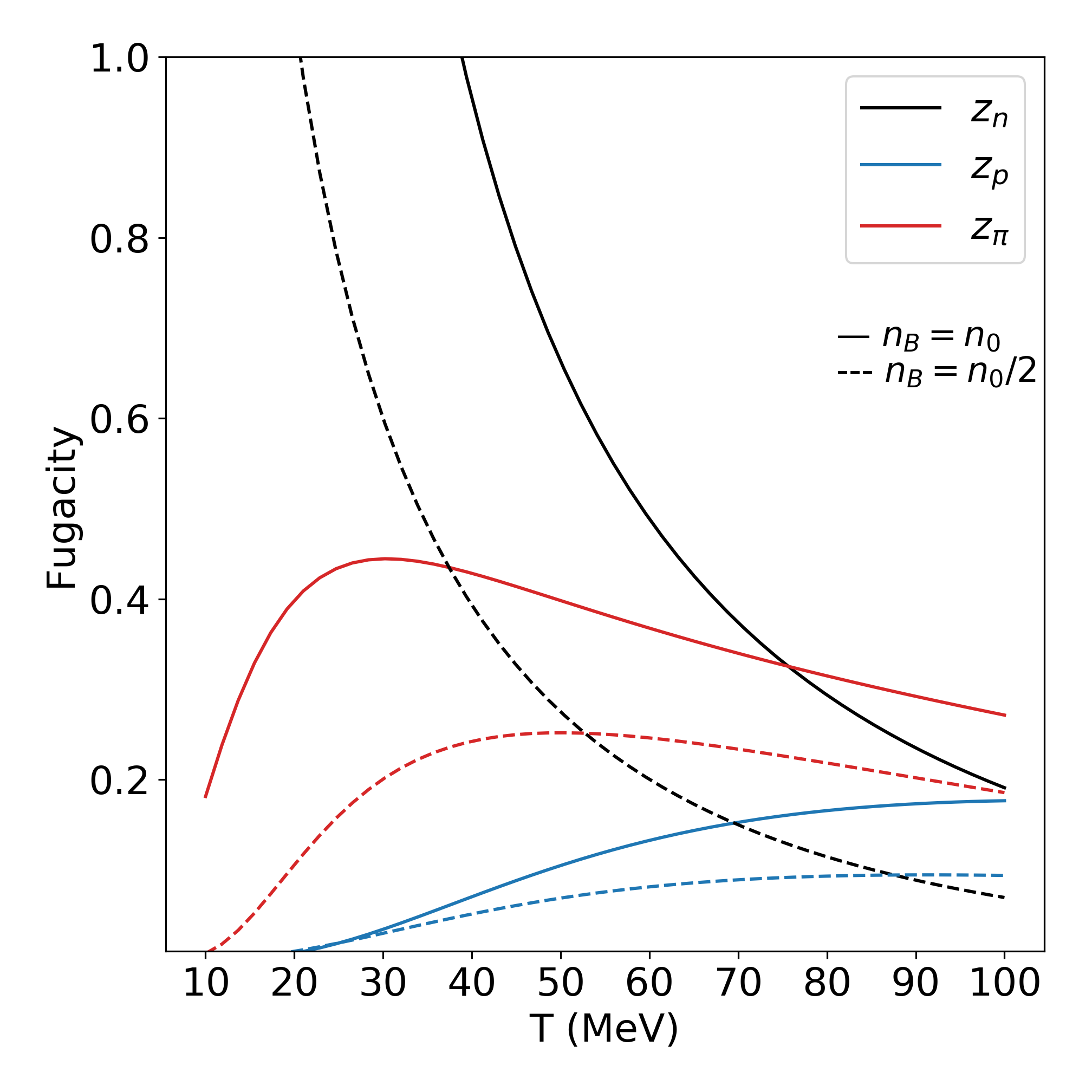}
\caption{Pion and nucleon fugacities in charge-neutral dense matter in $\beta$-equilibrium at $n_B=n_0$ (solid-curves) and $n_B=n_0/2$ (dashed-curves) are shown as function of temperature.}
\label{fig:fugacity}
\end{figure}

The fugacities of pions and nucleons at baryon density $n_B=n_0/2$ and $n_B=n_0$ as a function of the temperature are shown in Fig.~\ref{fig:fugacity}. 
It is interesting to note that $z_{\pi^-} $ and $z_p$ remain small over a wide range of temperatures. As expected in neutron-rich matter, the fugacity of neutrons is large, and the virial expansion for pion-neutron interactions is reliable only at high temperature. In what follows we consider matter at $n_B< 1.5 ~n_0$ and $T>25$ MeV and calculate the equation of state and weak interaction rates using the hybrid model in which pion-nucleon interactions are accounted for through the second-virial coefficient and nuclear interactions are treated in mean field theory.

\section{\label{sec:implications}Astrophysical Implications} 
\subsection{\label{sec:EOS} Equation of State} 
\begin{figure}[h]
\includegraphics[width=.45\textwidth]{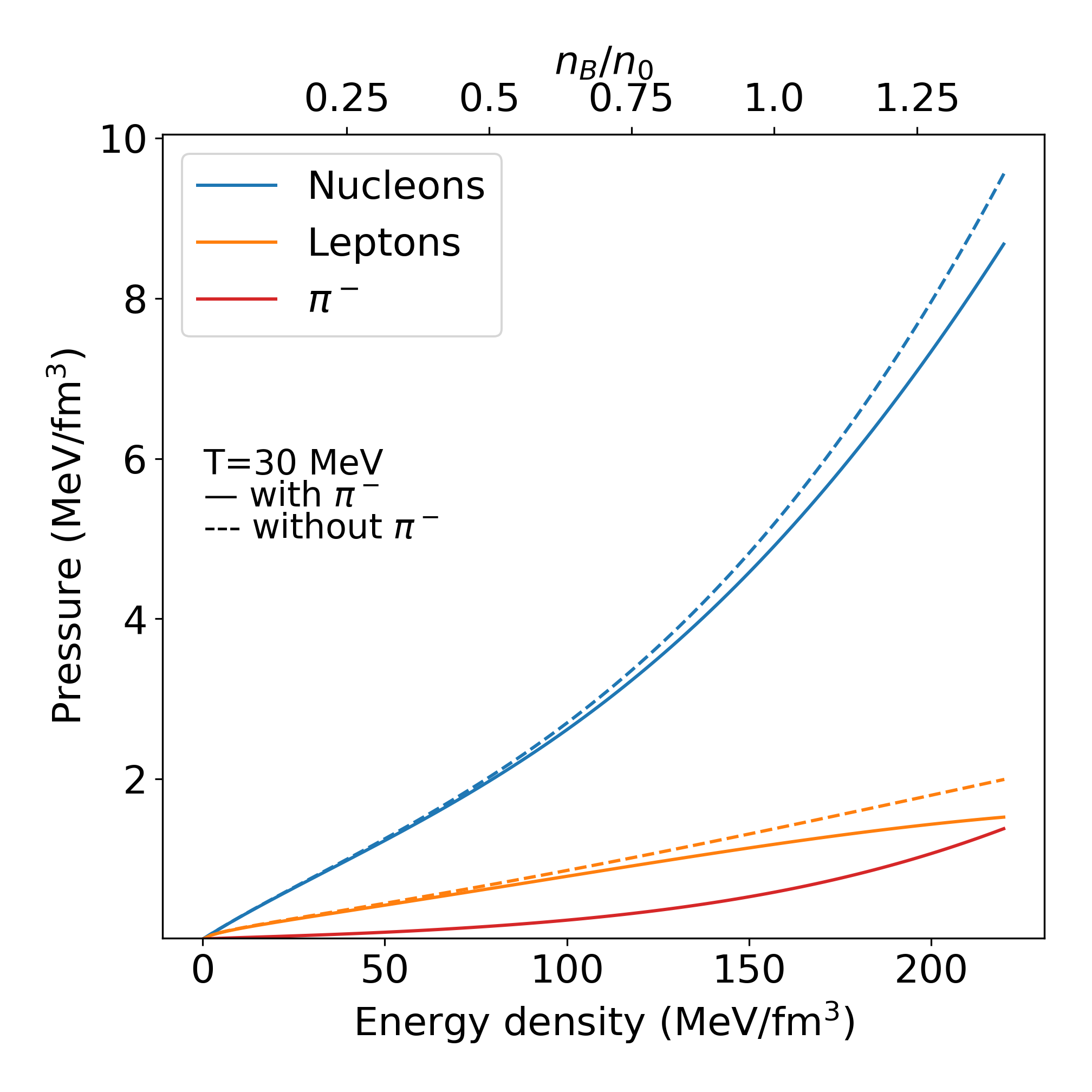}
\caption{The equation of state of hot dense matter with and without the inclusion of negative pions. The density axis corresponds to matter containing negative pions.}
\label{fig:EOSfixedT}
\end{figure}

At a given baryon density and temperature, pions alter the EOS in two ways. First, they make a small contribution to the pressure and energy density. The pion contribution to the pressure and the energy density, obtained in the relativistic virial expansion to leading order in the pion and nucleon fugacities, $z_{\pi^-}$ and $z_N$, are given by 
\begin{equation}
\begin{split}
P_{\pi^-} & = T z_{\pi^-} \int \frac{dk k^2}{2 \pi^2} ~\exp{ \left( -\beta \epsilon_\pi (k )\right)} \\
 & + T z_{\pi^-} \sum_{N=n,p} z_N b_2^{N\pi^-} \,,
 \end{split}
\label{eq:pvirial} 
\end{equation}
and
\begin{equation}
\begin{split}
\epsilon_{\pi^-} & = z_{\pi^-} \int \frac{dk k^2}{2 \pi^2} ~\epsilon_\pi(k) ~\exp{\left(-\beta \epsilon_\pi(k )\right)} \\
 & + z_{\pi^-} \sum_{N=n,p} z_N \frac{\partial b_2^{N\pi^-}}{\partial \beta } \,,
 \end{split}
\label{eq:evirial}
\end{equation}
respectively, where $\epsilon_\pi(k)= \sqrt{k^2+m^2_\pi}$ is the free pion dispersion relation \cite{Dashen:1969}. In Eqns.~\ref{eq:pvirial} \& \ref{eq:evirial}, the second line contains the contribution due to interactions between nucleons and pions. Second, by furnishing negative charge, pions increase the proton fraction, and decrease the lepton fraction. This has an important effect on the EOS, because any reduction in the asymmetry between neutron and protons lowers the pressure at fixed density. The individual contributions to the pressure at $T=30$ MeV are shown as a function of the energy density in Fig.~\ref{fig:EOSfixedT}. 
Here, the energy density and pressure of nucleons are calculated in mean field theory using Eq.~\ref{eq:enucleon}, and the leptons are treated as an ideal Fermi gas. From the figure we can infer that the dominant effect of pions is to alter the nucleon contribution. The symmetric state, with a higher proton fraction, is softer and has lower pressure at a given energy density.  

\begin{figure}[h]
\includegraphics[width=.45\textwidth]{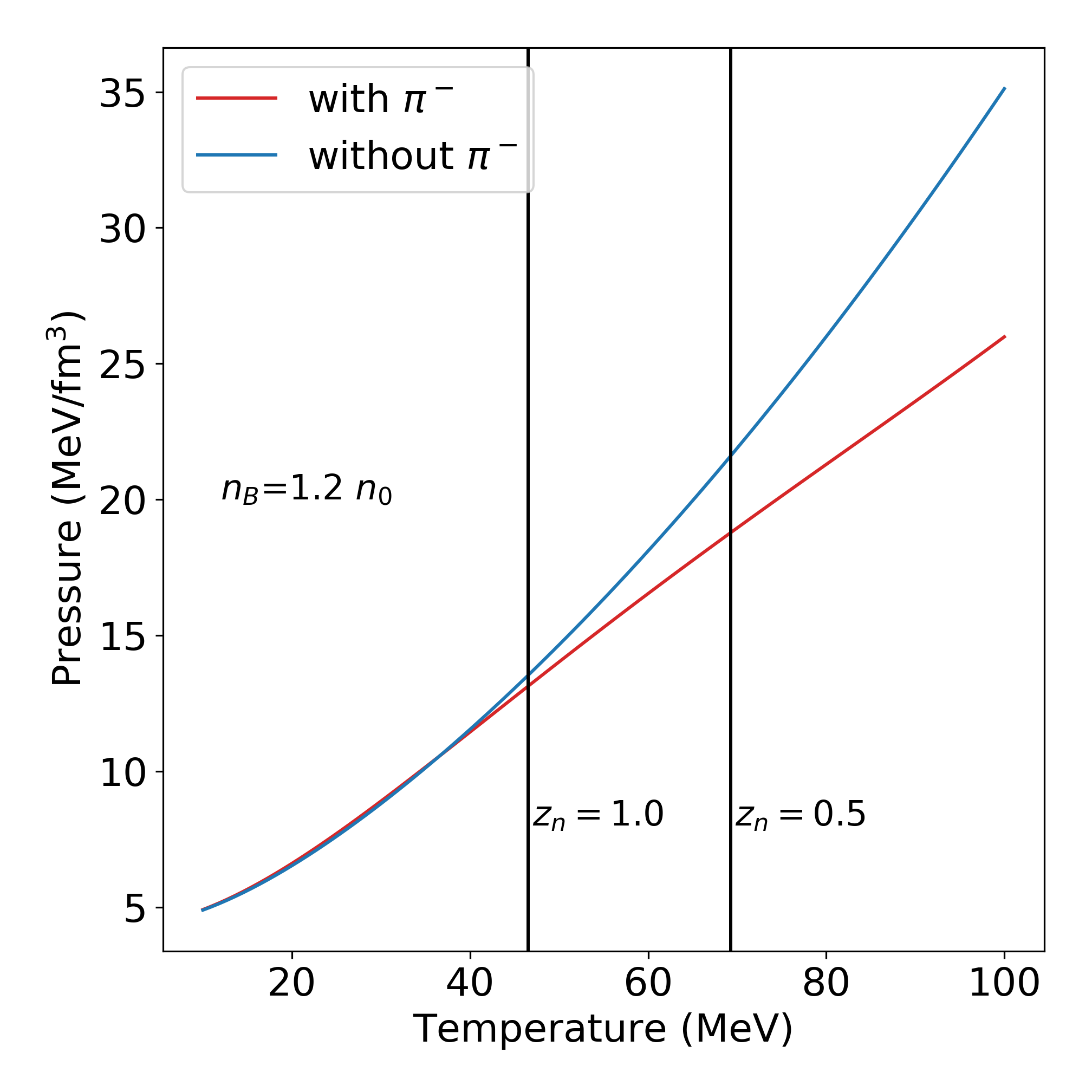}
\caption{The equation of state of hot dense matter with and without the inclusion of negative pions.}
\label{fig:EOS_T}
\end{figure}
Fig.~\ref{fig:EOS_T} shows the temperature dependence of the pressure at fixed baryon density. Motivated by recent simulations that suggest high temperatures, $T\simeq 20-50$ MeV, are realized in neutron star mergers even at baryon density $n_B \simeq n_0$ \cite{Perego:2019}, we chose a rather large range of $T$ to assess the temperatures at which pions would have the most dramatic effects on the EOS. From the figure, we see that pions play a role and soften the EOS when $T\gtrsim 40$ MeV. We find that at $n_B=1.2 n_0$ pions decrease the total pressure by about 9\% at $T=60$ MeV. Our study suggests that these effects can be significantly larger at higher density. However, methods beyond the virial expansion are needed to ascertain their importance.

\subsection{\label{sec:weak} Neutrino Mean Free Paths} 
The mean free path of neutrinos and anti-neutrinos in hot dense matter influences aspects of supernovae dynamics \cite{Janka:2012wk}, the observable signatures of neutrinos from proto-neutron stars \cite{Burrows:1986me,Pons:1998mm,Roberts:2017pns}, and is expected to play a role in neutron star mergers \cite{Sekiguchi:2011zd,Perego:2019}. At the densities and temperature encountered in these environments, all three flavors of neutrinos are produced, and contribute to the transport of energy, momentum, and lepton number. In matter containing nucleons and leptons, $\nu_e$ and $\bar{\nu}_e$ interact most strongly as they encounter both charged current and neutral current interactions with nucleons and leptons. The $\mu$ and $\tau$ neutrinos are coupled to matter only through their neutral-current interactions as their energies are not adequate to create the heavy charged leptons in the final state. Here, for the first time, we show that the presence of pions allows for new charged current reactions for muon neutrinos.  We find below that these reactions significantly reduce the $\nu_\mu$ and $\xbar{\nu}_\mu$ mean free paths. 

We find that the most important reactions are $\nu_\mu + \pi^- \rightarrow \mu^-$ and $\xbar{\nu}_\mu + \mu^- \rightarrow \pi^-$. The low energy effective Lagrangian that describes these weak processes is 
\begin{equation} 
{\cal L}=- \frac{G_F \cos{\theta_C}}{\sqrt{2}} f_\pi~\partial^\alpha \pi^- ~\bar{\psi}_{\nu_\mu} (\gamma_\alpha(1 -\gamma_5) )\psi_\mu\,,
\end{equation} 
where $f_\pi=130.4$ MeV is the pion decay constant \cite{Gasser:2010}. The amplitude-squared for the process $\xbar{\nu}_\mu + \mu^- \rightarrow \pi^-$ is obtained by summing over spin states of the muon in the initial state and is given by  
\begin{equation}
|A|_{\bar{\nu}_\mu}^2 = 2(G_F \cos{\theta_C f_\pi })^2 m_\mu^2 (E_\pi^2-p_\pi^2-m_\mu^2)\,, 
\end{equation}
where $E_\pi$ and $p_\pi$ are the pion energy and momentum, respectively, and $m_\mu$ is the mass of the muon.  
 In the vacuum, energy and momentum conservation forbids the process $\nu_\mu + \pi^- \rightarrow \mu^-$. However, in dense matter the modification of the pion dispersion relation, which we discuss in detail below, allows for this process when the pion momenta and energy satisfy $E_\pi^2-p_\pi^2 < m_\mu^2$. In this case, the amplitude-squared is obtained by summing over spin states of the muon in the final state and is given by  
\begin{equation}
|A|_{\nu_\mu}^2 = 2(G_F \cos{\theta_C} f_\pi)^2 m_\mu^2 (m^2_\mu - (E_\pi^2 - p_\pi^2)) \,. 
\end{equation}
We note that the amplitude-squared is proportional to the square of the lepton mass \textemdash{} a well-known fact that suppresses the decay of pions to electrons. It is for this reason that we focus on interactions involving only muon neutrinos in this work. 

Using Fermi's Golden rule, the mean free path of $\bar{\nu}_\mu$ due to the inverse decay reaction is given by 
\begin{equation}
\begin{split}
\frac{1}{\lambda_{\bar{\nu}_\mu}(E_{\bar{\nu}_\mu})} & = \int \frac{d^3\vec{p}_\mu}{(2\pi)^3 2 E_\mu} \int \frac{d^3\vec{p}_\pi}{(2\pi)^3 2 E_\pi} f_\mu (1+g_\pi) 
\\
& \times (2\pi)^4 \delta^4(P_\mu + P_{\bar{\nu}_\mu} - P_\pi) |A|_{\bar{\nu}_\mu}^2 \end{split}
\label{eq:numubar}
\end{equation}
where $g_\pi$ and $f_\mu$ are the Bose-Einstein distribution for pions and Fermi-Dirac distribution for muons, respectively. When kinematically allowed, the mean free path of $\nu_\mu$ due to the charged current reaction is 
\begin{equation}
\begin{split}
\frac{1}{\lambda_{\nu_\mu}(E_{\nu_\mu})} & = \int \frac{d^3\vec{p}_\mu}{(2\pi)^3 2 E_\mu} \int \frac{d^3\vec{p}_\pi}{(2\pi)^3 2 E_\pi} g_\pi (1-f_\mu) \\
& \times (2\pi)^4 \delta^4(P_\pi + P_{\nu_\mu} - P_\mu) |A|_{\nu_\mu}^2 \,.  
\end{split}
\label{eq:numu}
\end{equation}
The integrals appearing in Eqs.~\ref{eq:numubar} and \ref{eq:numu} can be further simplified and we find that 
\begin{eqnarray}
\frac{1}{\lambda_{\bar{\nu}_\mu}(E_{\bar{\nu}_\mu})} &=& \frac{1}{16 \pi E_\nu^2} \int^{p_h} _{p_l} dp_\pi \frac{p_\pi}{E_\pi} f_\mu (1+g_\pi) |A|^2_{\bar{\nu}_\mu}\,,\\
\frac{1}{\lambda_{\nu_\mu}(E_{\nu_\mu})} &=& \frac{1}{16 \pi E_\nu^2} \int^{p_h} _{p_l} dp_\pi \frac{p_\pi}{E_\pi} g_\pi (1-f_\mu) |A|^2_{\nu_\mu}\,. 
\end{eqnarray}
The limits of the pion momentum integral, $p_l$ and $p_h$, arise due to the energy conservation. For the $\xbar{\nu}_\mu + \mu^- \rightarrow \pi^-$ reaction, the limits are determined to ensure that  
\begin{equation}
-1 \le \frac{E_\pi}{p_\pi} - \frac{E_\pi^2-p_\pi^2-m_\mu^2}{2 p_\pi E_\nu} \le 1\,, 
\label{eq:constraint_inverse_pi_decay}
\end{equation}
and for the $\nu_\mu + \pi^- \rightarrow \mu^-$ reaction, they are obtained to ensure that 
\begin{equation}
-1 \le \frac{E_\pi}{p_\pi} + \frac{E_\pi^2-p_\pi^2-m_\mu^2}{2 p_\pi E_\nu} \le 1\,. 
\label{eq:constraint_inverse_mu_decay}
\end{equation}
When $E_\pi > p_\pi$, Eq.~\ref{eq:constraint_inverse_pi_decay} can be satisfied when $E_\pi^2-p_\pi^2 \ge m_\mu^2$ and Eq.~\ref{eq:constraint_inverse_mu_decay} can be satisfied when $E_\pi^2-p_\pi^2 \le m_\mu^2$. For example, at $p_\pi = 125$ MeV, which is near the typical momentum for a pion, at nuclear density and $T=30$ MeV the inverse pion decay reaction is allowed for neutrinos with energies approximately between 8 MeV and 45 MeV. At high momenta, $p_\pi \gtrsim 200$ MeV, we find that our dispersion relation allows for $E_\pi < p_\pi$ and in this case both reactions are allowed and the range of allowed neutrino energies is only bounded from below. For the inverse pion decay reaction this lower bound is at very high neutrino energies, but for the $\nu_\mu + \pi^- \rightarrow \mu^-$ reaction the lower bound is around 20-40 MeV in matter at nuclear density and $T=30$ MeV.

To calculate the neutrino mean free paths we need the pion dispersion relation to determine the relationship between the pion energy and momentum in matter. In general, this is given by 
\begin{equation} 
E_{\pi^-}(p) = \sqrt{p^2+m_\pi^2} + \Sigma_{\pi^-}(p)\,, 
\label{eq:pion_dispersion}
\end{equation} 
where $ \Sigma_{\pi^-}(p)$ is the self-energy. 
\begin{figure}[h]
\includegraphics[width=.45\textwidth]{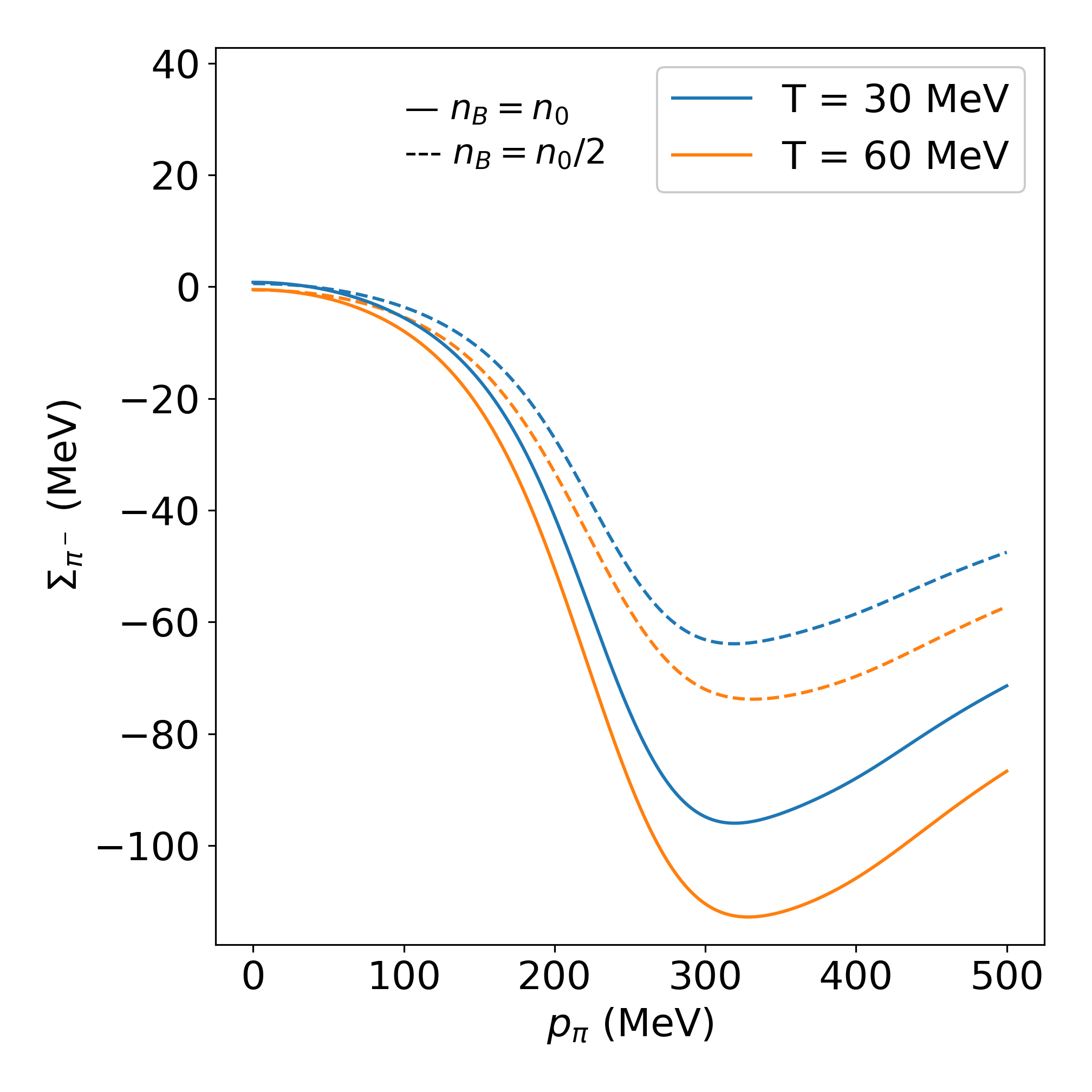}
\caption{Pion self energy predicted by our model at a few representative temperatures and baryon densities.}
\label{fig:pi_self_energy}
\end{figure}
We propose a simple model to calculate the real part of $ \Sigma_{\pi^-}(p)$ using the one-loop approximation. The model is constructed to be consistent with the predictions of the virial expansion and the pion-self energy is given by 
\begin{equation} 
\Sigma_{\pi^-}(p) = \int \frac{d^3k}{(2\pi)^3}~\sum_{N=n,p} f_N(E_N(k)) ~V^{ps}_{N\pi^-}(p_{cm}) \,,
\label{eq:sigma_pp}
\end{equation} 
where the pion-nucleon interaction is directly proportional to the phase shifts 
\be 
V^{(ps)}_{N\pi^-} (p_{\rm cm})= - \alpha \sum_{I,l,\nu} (2l+1) \frac{ 2\pi\delta_{l,\nu}^{I}}{ {\bar m} ~p_{\rm cm}} \,. 
\label{eq:psudopotential}
\ee 
Here, $p_{\rm cm} = \bar m \sqrt{\frac{p^2}{m^2_\pi}+\frac{k^2}{m^2_N}-  \frac{2 p k}{m_\pi m_N} \cos{\theta}}$ is the center of mass momentum, and $\bar{m}=m_\pi m_N/(m_N+m_\pi)$ is the reduced mass. The sum is over allowed values of the isospin ($I$), angular momentum values ($l$), and nucleon spin-projections ($\nu=+,-$). Note, that the pseudo-potential is proportional to $\delta_{l,\nu}^{I}$ and differs from the other choices such as the $T-$matrix which is proportional to  $\sin{\delta_{l,\nu}^{I}}$ or the $R-$ matrix that is proportional to $\tan{\delta_{l,\nu}^{I}}$. This choice for the pseudo-potential is motivated by the observation that the second-virial coefficient is also  proportional to $\delta_{l,\nu}^{I}$. In addition, Fumi's theorem -- a well-known result in condensed matter physics shows that the calculation of the ground state energy shift due to interactions between particles in a gas and an impurity can be obtained if the pseudo-potential of the form in Eq.~\ref{eq:psudopotential} is used as an effective interaction \cite{Mahan:2000}.

A fudge factor $\alpha$ is introduced to ensure that the number density we obtain using this dispersion relation matches the result in Eq.~\ref{eq:pion_number} obtained in the virial expansion. We find that the $\pi^-$ interaction with neutrons dominates the self-energy, and in what follows we neglect the contribution due to $\pi^-$ proton interactions. The self-energy obtained in this way is shown in Fig.~\ref{fig:pi_self_energy}. We employ the experimentally measured phase shifts up to $p_{\rm cm} \approx 350$ MeV and assume that it remains constant at higher momentum. The values for the fudge factor $\alpha$ used to ensure consistency with the virial result are given in table \ref{tb:alpha}.  
\begin{table}[h]
\begin{tabular}{|c|c|c|}
\hline
 $\alpha$   & T=30 MeV & T=60 MeV \\ \hline
 $n_B=0.5n_0$ & 0.183 & 0.216 \\ \hline
 $n_B=1.0n_0$ & 0.139 & 0.171 \\ \hline
\end{tabular}
\caption{Values of the fudge factor $\alpha$ needed to obtain consistency.}
\label{tb:alpha}
\end{table}
It is interesting to note that the fudge factor $\alpha \simeq 1/(2\pi)$ but we do not have an explanation for why this is the case. 

Although our model for $\Sigma_{\pi^-}(p)$ is admittedly very crude, the modest variation of $\alpha$ over a broad range of densities and temperatures is reassuring. It suggests that our ansatz for the pseudo-potential provides a fair description of the momentum dependence of pion-nucleon interactions (we have explicitly checked that other choices such as the $T-$matrix which is proportional to  $\sin{(\delta_{l,\nu}^{I})}$ would produce a larger variation of  $\alpha$ with temperature and density). We have examined the general behaviour of the pion dispersion relation we obtain and find that it is physically plausible. The substantial reduction in the pion energy seen in Fig.~\ref{fig:pi_self_energy} at $p_\pi \simeq 300$ MeV is due to the strong p-wave attractive interaction, and the small increase at $p=0$ arises to weak and repulsive s-wave interaction. The group velocity of the pions is also roughly consistent with general expectations. It is small at low momentum and approaches $c$ (speed of light) at large momenta. At intermediate values $\simeq 350$ MeV we find that the model predicts a group velocity that can exceed $c$ by a few percents - a mild deficiency given the approximations of our model. First, the pseudo-potential in Eq. \ref{eq:psudopotential} was employed in the Born approximation to calculate $\Sigma_{\pi^-}$ and it provided a direct relationship between the self-energy and the phase shifts in Eq.~\ref{eq:sigma_pp}. This relationship is exact only in the limit when one can neglect correlations between nucleons and nucleon recoils \cite{Mahan:2000}. Second, our approximation that the phase shift-remains constant for $p_{\rm cm} \gtrsim 350$ MeV, has an effect on the behaviour of the pion self-energy at these large momenta. Third, we have neglected the imaginary part of the pion self-energy in the matter. The imaginary part arises due to two-loop contributions involving two nucleons in the medium. For these reasons, we view our model as the first step towards more realistic calculations. 

\begin{figure}[h]
\includegraphics[width=.45\textwidth]{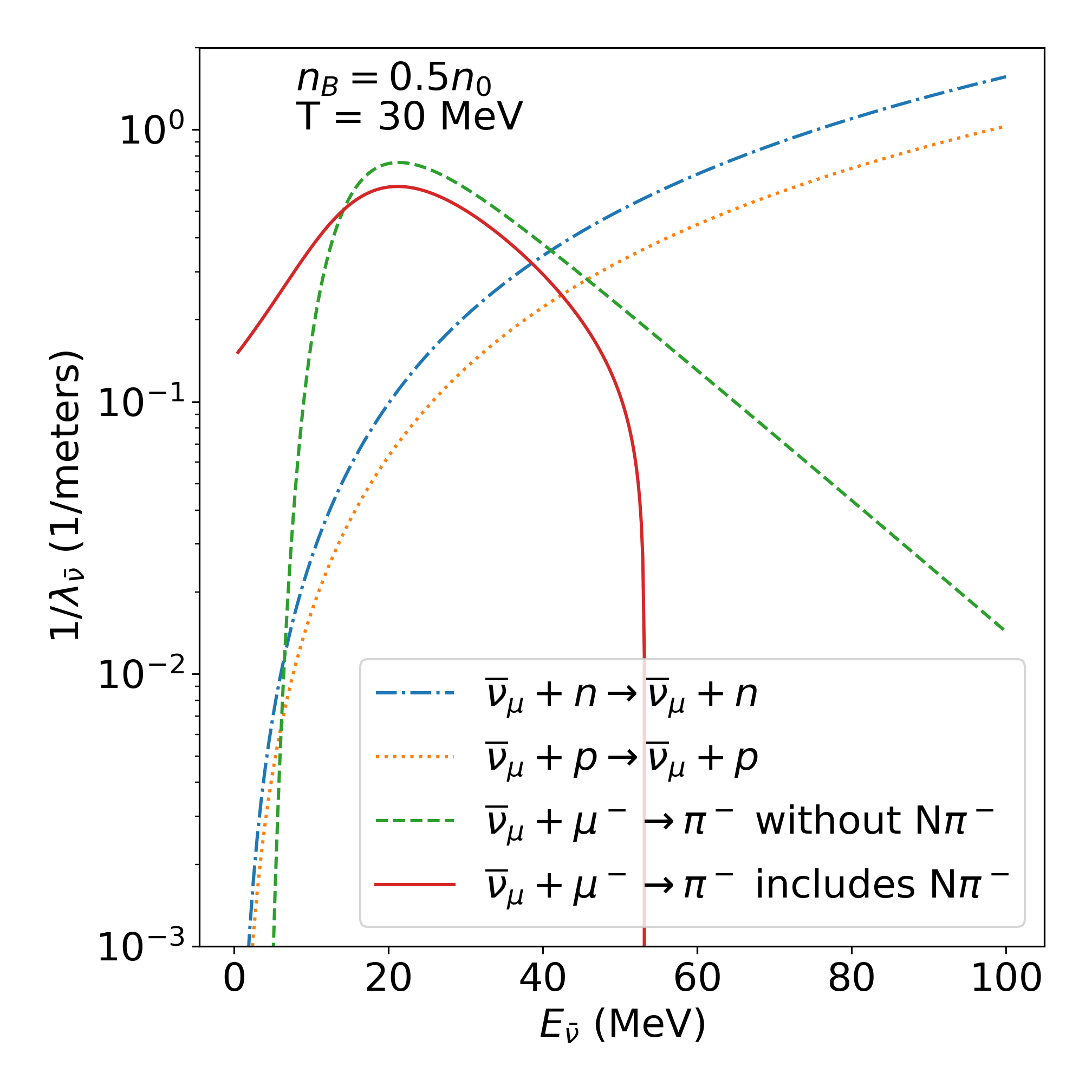}
\caption{Antineutrino inverse mean free paths due to the inverse pion decay reaction, with and without $N\pi^-$ interactions included, are compared to the neutral current reactions involving nucleons.}
\label{fig:nubar_mfp}
\end{figure}
The inverse mean free path due to the reaction $\xbar{\nu}_\mu + \mu^- \rightarrow \pi^-$ in matter containing pions at $n_B=0.5~n_0$ and $T=30$ MeV is shown in Fig.~\ref{fig:nubar_mfp}. The dashed-green curve is calculated using the vacuum dispersion relation for the pions. Since pions only appear in the final state, this curve depends only weakly on the model for pion-nucleon interactions. The solid-red curve is obtained using the dispersion relation in Eq.~\ref{eq:pion_dispersion}, and the self-energy depicted in Fig.~\ref{fig:pi_self_energy}. Here we see the strong influence of the in-medium dispersion relation, especially at large neutrino energy. The reduction in the pion energy due to its large and attractive p-wave interaction with nucleons implies that a large momentum pion in the final state is unable to satisfy energy and momentum conservation in the medium. The rapid decrease in the inverse mean free path depicted by the solid-red curve reflects these severe kinematic constraints. At lower neutrino energy, the in-medium dispersion relation leads to a significant reduction of the $\xbar{\nu}_\mu$ mean free path. It is remarkable that at these low energies neutrino processes involving a sparse population of muons and pions are significantly more important than processes involving nucleons and electrons. Neutral current reactions $\xbar{\nu}_\mu + X \rightarrow \xbar{\nu}_\mu + X$ where $X=n,p,e^-$ have been studied extensively in earlier work \cite{Horowitz:1990it,Reddy:1998hb} and we use the open-source computer codes from the neutrino opacity library, nuOpac\cite{Roberts:2017}, to calculate the neutrino mean free paths. The contributions from the reactions $\xbar{\nu}_\mu + n\rightarrow \xbar{\nu}_\mu + n$, and $\xbar{\nu}_\mu + p\rightarrow \xbar{\nu}_\mu + p$ are shown as the blue dot-dashed and orange dotted curves in Fig.~\ref{fig:nubar_mfp}. Neutral reactions involving electrons, not shown in the figure, are smaller than that due to the nucleons. 

\begin{figure}[h]
\includegraphics[width=.45\textwidth]{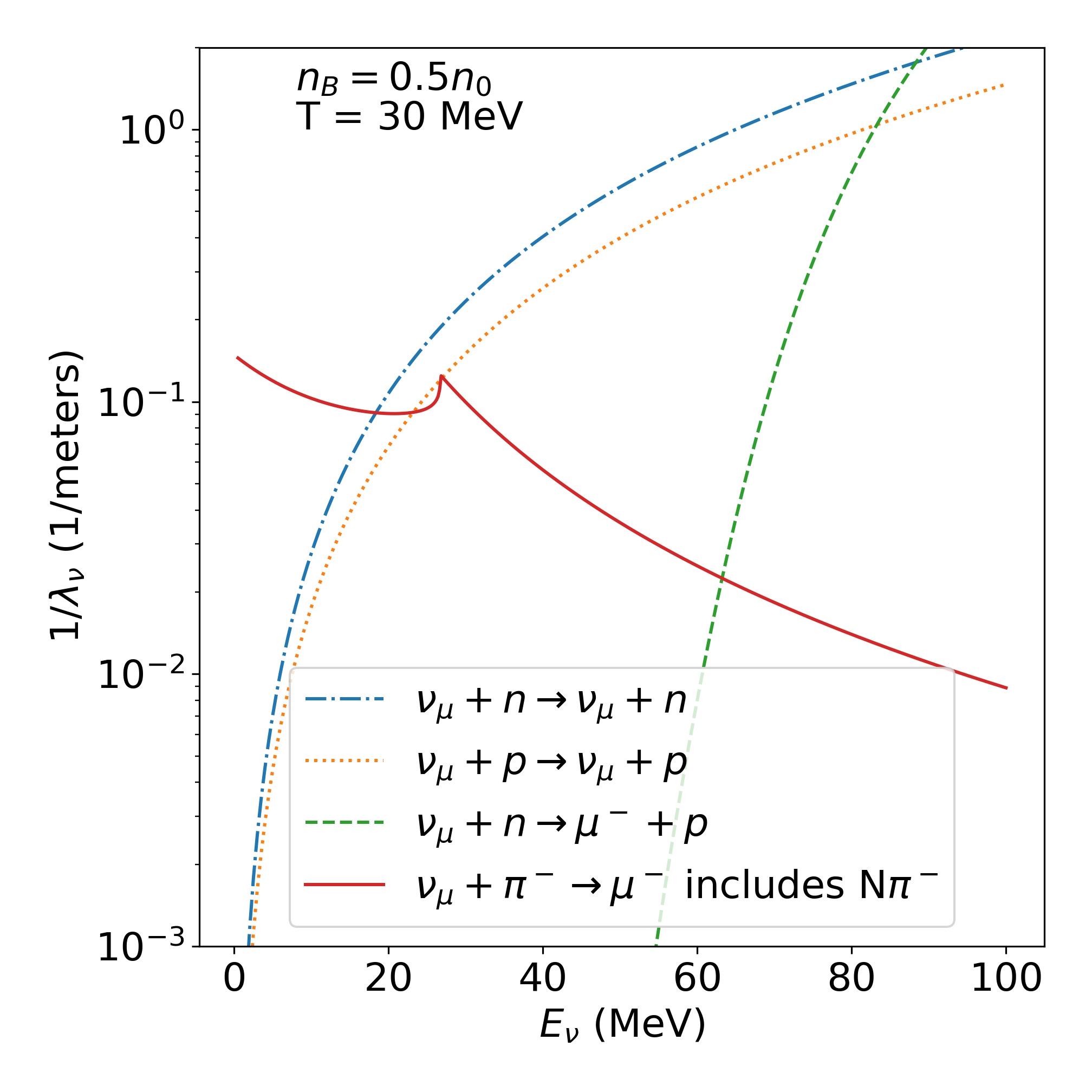}
\caption{Neutrino inverse mean free path due to the inverse muon decay reaction is compared with the mean free path due to neutral and charged current reactions involving nucleons.}
\label{fig:nu_mfp}
\end{figure}
The mean free path of muon neutrinos in matter containing pions at $n_B=0.5~n_0$ and $T=30$ MeV is shown in Fig.~\ref{fig:nu_mfp}. This process, which is forbidden in the vacuum, is sensitive to the pion dispersion relation and their abundance. Again, the result, depicted by the solid-red curve, shows some remarkable features. At low energy, the process involving pions is dominant. It remains more important than the charged current reactions involving nucleons shown as the dashed-green curve even at higher energies. Neutral current scattering off nucleons, shown by the blue dot-dashed and orange dotted curves, continues to be the dominant reaction for thermal neutrinos under these specific conditions. The sharp feature in the solid-red curve at $E_\nu \simeq 30$ MeV is due to the non-monotonic behaviour of the kinematic constraint in Eq. \ref{eq:constraint_inverse_mu_decay}.

\subsection{\label{sec:eqbulk}Weak equilibration and bulk viscosity}
Bulk viscosity offers a mechanism to damp density oscillations in matter and plays a role in neutron star dynamics \cite{Cutler:1990}. For example, dissipative effects in neutron star mergers influence the lifetime of the hot dense hyper-massive neutron star and the post-merger gravitational-wave emission\cite{Alford:2017rxf}. Bulk viscosity arises due to non-equilibrium reactions that convert chemical energy into thermal energy. This conversion happens because the equilibrium chemical composition of matter changes with density; therefore, the density perturbations induce inelastic reactions.

In neutron stars, where relevant dynamical timescales are of the order of milliseconds, weak reactions play the dominant role in determining the bulk viscosity \cite{Sawyer:1989}. In dense nuclear matter, the reaction $e^- +p \leftrightarrow n + \nu_e$, often referred to as the URCA reactions in astrophysics, and the modified URCA reaction $e^- +p +n \leftrightarrow n + n + \nu_e$ change the proton fraction when perturbed and are generally considered to be the main source of bulk viscosity. Recent work have investigated the role of these weak reactions involving nucleons in dense matter with and without neutrino trapping at high temperatures \cite{Alford:2019qtm,Alford:2019kdw}. At $n_B=0.5 n_0$ and $T=30$ MeV the results in Ref.~\cite{Alford:2019kdw} indicate that the beta-equilibrium relaxation time for these reactions is about $10^{-7}$ s for the neutrino free case, and about $10^{-9}$ s when neutrinos are trapped. However, we expect that under similar conditions reactions involving pions and nucleons, would proceed on much faster timescales due to the strong interaction, allowing for faster equilibration of the proton fraction. 

Consider a density perturbation in which the final equilibrium state contains a larger neutron fraction. In the absence of reactions involving pions, electron capture reactions $e^- +p \rightarrow n + \nu_e$ and $e^- +p +n \rightarrow n + n + \nu_e$ generate the needed neutrons. When these reactions are out of equilibrium they generate heat and dissipation. In the presence of pions and muons there are additional reaction channels that can play a role. These include  
\begin{eqnarray} 
\pi^- + p + n &\leftrightarrow& n + n\,, \label{eq:nucleon_bulk_viscosity_reaction} \\ 
\mu^- &\leftrightarrow& \pi^- + \nu_\mu\,,\\
\mu^- &\leftrightarrow& e^- + \bar{\nu}_{e} + \nu_\mu \,,\\
\pi^- &\leftrightarrow& \mu^- + \bar{\nu}_\mu\,. 
\end{eqnarray} 
The non-leptonic reactions mediated by the strong interaction proceed on a timescale that is much faster than the weak reactions involving leptons. Consequently, the timescale for equilibration is set by the weak reactions involving pions. These observations suggest that the role of weak interactions involving nucleons will be a sub-dominant process when pions are present. A detailed analysis of these reaction pathways, and their role in determining the bulk viscosity warrants more work and will reported elsewhere \cite{pionbulkviscoisity}. We note that although the modification to the pion dispersion relation allows for the process $ \pi^-+ p \rightarrow n$ which is forbidden in the vacuum, in practice we find that these reactions can occur only when the nucleon momentum is very large. At $T=30$ MeV and $n_B=n_0$ the minimum nucleon momentum needed for this process is $\simeq 730$ MeV. Since this is much larger than the momentum of thermal nucleons, $p_{\rm nuc} \simeq \sqrt{3 MT} \simeq 290$ MeV we expect that its contribution will be negligible. 

\section{\label{sec:conclusion}Conclusion}
Our study shows that negatively charged pions are an important degree of freedom in hot dense matter encountered in astrophysics. The virial expansion provides a model-independent approach to include pion-nucleon interactions when the fugacities are small and provides strong evidence for the enhancement of the pion number density  due to pion-nucleon interactions.  At densities and temperatures relevant for the study of neutron star mergers and core-collapse supernovae, the proton and pion fugacity are small, but the neutron fugacity can be large. To describe such matter we proposed a simple hybrid model that provides qualitative insights about the role of pions in hot neutron-rich matter. The attractive p-wave interaction between nucleons and pions was found to significantly increase the density of pions and the proton fraction in the charge-neutral matter in $\beta-$equilibrium. At a baryon density $n_B\simeq n_0$, pion contributions to the thermodynamics become relevant when $T> 25$ MeV. At $T=60$ MeV $n_B\simeq n_0$, we find that pions lower the pressure by about $10\%$, and at $T=100$ MeV they lower the pressure by about $30\%$. A naive extrapolation to $n_B = 2 n_0$, suggests that pions could have dramatic effects on the EOS, and the transport properties even at low temperature. However, further work is needed to study matter at these higher densities where the virial approximation fails, and pion condensation becomes a possibility. 

Our most significant finding is that even a relatively small number fraction of pions alters the neutrino mean free paths and the reaction pathways for equilibration of the proton and lepton fractions. The presence of pions and muons allows for additional reactions such as $\nu_\mu + \pi^- \rightarrow \mu^-$ and $\bar{\nu}_\mu + \mu^- \rightarrow \pi^-$. These reactions make the dominant contribution to the mean free path of low energy muon neutrinos. Charged current interactions are efficient at transferring energy; thus, the shorter mean free paths of the muon neutrinos should impact energy transport in protoneutron stars and neutron star mergers.  

We believe that our results establish the need to include pions as explicit degrees of freedom in the calculations of the EOS and transport properties of hot dense matter encountered in astrophysics. However, the approximations made in this study, and discussed extensively in previous sections, warrant a critical assessment. In particular, the pion-dispersion relation in the dense medium should be calculated in a microscopic theory in which pion-nucleon and nucleon-nucleon interactions are treated consistently. Extending calculations of hot dense matter based on Chiral perturbation theory \cite{Wellenhofer:2015qba} to include dynamical pions will be ideally suited for this purpose. 

Finally, we remark on the need to revisit the possibility of s-wave $\pi^-$ condensation in neutron star matter. Although we were deliberate in restricting our analysis to high temperature and relatively low-density in this study to ensure that we had a model-independent basis for our claims, our calculations suggest that pion-nucleon interactions are weak for low-momentum pions. From Fig.~\ref{fig:pi_self_energy} we can deduce that the energy-shift of a zero-momentum $\pi^-$ in neutron-rich matter at nuclear density is negligible. We believe that it is likely to remain so even at higher density and lower temperature. This would imply that threshold for s-wave $\pi^-$ condensation would be $\mu_e \gtrsim m_\pi^-$, and that s-wave $\pi^-$ condensation will occur at $n_B\le 2n_0$. A detailed study of $\pi^-$ condensation and its implications for neutron stars will be presented elsewhere.   
\section*{Acknowledgments} 
We thank Martin Hoferichter for useful discussions and for providing pion-nucleon phase-shift data. We thank Mark Alford, Steven Harris, Madappa Prakash and Corbinian Wellenhofer for reading manuscript and for their comments and suggestions. B. F. thanks Sam Kowash, Tyler Blanton, and Meredith Fore for useful discussions and support. B. F. acknowledges support from the SciDAC Grant No. A18-0354-S002 (DE-SC0018232). We would also to thank an anonymous referee for several useful suggestions and for noting that we had missed a factor of $2\pi$ in defining the pseudo-potential in an earlier version of the paper. The work of S. R. was supported by the U.S. DOE under Grant No. DE-FG02- 00ER41132. B. F. also acknowledges supported by the National Science Foundation Graduate Research Fellowship Program under Grant No. (DGE-1762114). 

\bibliography{main}

\begin{thebibliography}{40}%
\makeatletter
\providecommand \@ifxundefined [1]{%
 \@ifx{#1\undefined}
}%
\providecommand \@ifnum [1]{%
 \ifnum #1\expandafter \@firstoftwo
 \else \expandafter \@secondoftwo
 \fi
}%
\providecommand \@ifx [1]{%
 \ifx #1\expandafter \@firstoftwo
 \else \expandafter \@secondoftwo
 \fi
}%
\providecommand \natexlab [1]{#1}%
\providecommand \enquote  [1]{``#1''}%
\providecommand \bibnamefont  [1]{#1}%
\providecommand \bibfnamefont [1]{#1}%
\providecommand \citenamefont [1]{#1}%
\providecommand \href@noop [0]{\@secondoftwo}%
\providecommand \href [0]{\begingroup \@sanitize@url \@href}%
\providecommand \@href[1]{\@@startlink{#1}\@@href}%
\providecommand \@@href[1]{\endgroup#1\@@endlink}%
\providecommand \@sanitize@url [0]{\catcode `\\12\catcode `\$12\catcode
  `\&12\catcode `\#12\catcode `\^12\catcode `\_12\catcode `\%12\relax}%
\providecommand \@@startlink[1]{}%
\providecommand \@@endlink[0]{}%
\providecommand \url  [0]{\begingroup\@sanitize@url \@url }%
\providecommand \@url [1]{\endgroup\@href {#1}{\urlprefix }}%
\providecommand \urlprefix  [0]{URL }%
\providecommand \Eprint [0]{\href }%
\providecommand \doibase [0]{http://dx.doi.org/}%
\providecommand \selectlanguage [0]{\@gobble}%
\providecommand \bibinfo  [0]{\@secondoftwo}%
\providecommand \bibfield  [0]{\@secondoftwo}%
\providecommand \translation [1]{[#1]}%
\providecommand \BibitemOpen [0]{}%
\providecommand \bibitemStop [0]{}%
\providecommand \bibitemNoStop [0]{.\EOS\space}%
\providecommand \EOS [0]{\spacefactor3000\relax}%
\providecommand \BibitemShut  [1]{\csname bibitem#1\endcsname}%
\let\auto@bib@innerbib\@empty
\bibitem [{\citenamefont {Oertel}\ \emph {et~al.}(2017)\citenamefont {Oertel},
  \citenamefont {Hempel}, \citenamefont {Kl{\"a}hn},\ and\ \citenamefont
  {Typel}}]{Oertel:2016bki}%
  \BibitemOpen
  \bibfield  {author} {\bibinfo {author} {\bibfnamefont {M.}~\bibnamefont
  {Oertel}}, \bibinfo {author} {\bibfnamefont {M.}~\bibnamefont {Hempel}},
  \bibinfo {author} {\bibfnamefont {T.}~\bibnamefont {Kl{\"a}hn}}, \ and\
  \bibinfo {author} {\bibfnamefont {S.}~\bibnamefont {Typel}},\ }\href
  {\doibase 10.1103/RevModPhys.89.015007} {\bibfield  {journal} {\bibinfo
  {journal} {Rev. Mod. Phys.}\ }\textbf {\bibinfo {volume} {89}},\ \bibinfo
  {pages} {015007} (\bibinfo {year} {2017})},\ \Eprint
  {http://arxiv.org/abs/1610.03361} {arXiv:1610.03361 [astro-ph.HE]}
  \BibitemShut {NoStop}%
\bibitem [{\citenamefont {Bollig}\ \emph {et~al.}(2017)\citenamefont {Bollig},
  \citenamefont {Janka}, \citenamefont {Lohs}, \citenamefont
  {Mart\'{\i}nez-Pinedo}, \citenamefont {Horowitz},\ and\ \citenamefont
  {Melson}}]{muon_nova}%
  \BibitemOpen
  \bibfield  {author} {\bibinfo {author} {\bibfnamefont {R.}~\bibnamefont
  {Bollig}}, \bibinfo {author} {\bibfnamefont {H.-T.}\ \bibnamefont {Janka}},
  \bibinfo {author} {\bibfnamefont {A.}~\bibnamefont {Lohs}}, \bibinfo {author}
  {\bibfnamefont {G.}~\bibnamefont {Mart\'{\i}nez-Pinedo}}, \bibinfo {author}
  {\bibfnamefont {C.~J.}\ \bibnamefont {Horowitz}}, \ and\ \bibinfo {author}
  {\bibfnamefont {T.}~\bibnamefont {Melson}},\ }\href {\doibase
  10.1103/PhysRevLett.119.242702} {\bibfield  {journal} {\bibinfo  {journal}
  {Phys. Rev. Lett.}\ }\textbf {\bibinfo {volume} {119}},\ \bibinfo {pages}
  {242702} (\bibinfo {year} {2017})}\BibitemShut {NoStop}%
\bibitem [{\citenamefont {Migdal}(1973)}]{Migdal:1973}%
  \BibitemOpen
  \bibfield  {author} {\bibinfo {author} {\bibfnamefont {A.}~\bibnamefont
  {Migdal}},\ }\href {\doibase https://doi.org/10.1016/0370-2693(73)90640-0}
  {\bibfield  {journal} {\bibinfo  {journal} {Physics Letters B}\ }\textbf
  {\bibinfo {volume} {45}},\ \bibinfo {pages} {448 } (\bibinfo {year}
  {1973})}\BibitemShut {NoStop}%
\bibitem [{\citenamefont {Sawyer}(1972)}]{Sawyer:1972}%
  \BibitemOpen
  \bibfield  {author} {\bibinfo {author} {\bibfnamefont {R.~F.}\ \bibnamefont
  {Sawyer}},\ }\href {\doibase 10.1103/PhysRevLett.29.382} {\bibfield
  {journal} {\bibinfo  {journal} {Phys. Rev. Lett.}\ }\textbf {\bibinfo
  {volume} {29}},\ \bibinfo {pages} {382} (\bibinfo {year} {1972})}\BibitemShut
  {NoStop}%
\bibitem [{\citenamefont {{Migdal}}(1978)}]{Migdal:1978}%
  \BibitemOpen
  \bibfield  {author} {\bibinfo {author} {\bibfnamefont {A.~B.}\ \bibnamefont
  {{Migdal}}},\ }\href {\doibase 10.1103/RevModPhys.50.107} {\bibfield
  {journal} {\bibinfo  {journal} {Reviews of Modern Physics}\ }\textbf
  {\bibinfo {volume} {50}},\ \bibinfo {pages} {107} (\bibinfo {year}
  {1978})}\BibitemShut {NoStop}%
\bibitem [{\citenamefont {Weise}\ and\ \citenamefont
  {Brown}(1975)}]{Weise:1975}%
  \BibitemOpen
  \bibfield  {author} {\bibinfo {author} {\bibfnamefont {W.}~\bibnamefont
  {Weise}}\ and\ \bibinfo {author} {\bibfnamefont {G.}~\bibnamefont {Brown}},\
  }\href {\doibase https://doi.org/10.1016/0370-2693(75)90658-9} {\bibfield
  {journal} {\bibinfo  {journal} {Physics Letters B}\ }\textbf {\bibinfo
  {volume} {58}},\ \bibinfo {pages} {300 } (\bibinfo {year}
  {1975})}\BibitemShut {NoStop}%
\bibitem [{\citenamefont {Backman}\ and\ \citenamefont
  {Weise}(1975)}]{Backman:1975}%
  \BibitemOpen
  \bibfield  {author} {\bibinfo {author} {\bibfnamefont {S.-O.}\ \bibnamefont
  {Backman}}\ and\ \bibinfo {author} {\bibfnamefont {W.}~\bibnamefont
  {Weise}},\ }\href {\doibase https://doi.org/10.1016/0370-2693(75)90171-9}
  {\bibfield  {journal} {\bibinfo  {journal} {Physics Letters B}\ }\textbf
  {\bibinfo {volume} {55}},\ \bibinfo {pages} {1 } (\bibinfo {year}
  {1975})}\BibitemShut {NoStop}%
\bibitem [{\citenamefont {Migdal}\ \emph {et~al.}(1990)\citenamefont {Migdal},
  \citenamefont {Saperstein}, \citenamefont {Troitsky},\ and\ \citenamefont
  {Voskresensky}}]{Migdal:1990}%
  \BibitemOpen
  \bibfield  {author} {\bibinfo {author} {\bibfnamefont {A.}~\bibnamefont
  {Migdal}}, \bibinfo {author} {\bibfnamefont {E.}~\bibnamefont {Saperstein}},
  \bibinfo {author} {\bibfnamefont {M.}~\bibnamefont {Troitsky}}, \ and\
  \bibinfo {author} {\bibfnamefont {D.}~\bibnamefont {Voskresensky}},\ }\href
  {\doibase https://doi.org/10.1016/0370-1573(90)90132-L} {\bibfield  {journal}
  {\bibinfo  {journal} {Physics Reports}\ }\textbf {\bibinfo {volume} {192}},\
  \bibinfo {pages} {179 } (\bibinfo {year} {1990})}\BibitemShut {NoStop}%
\bibitem [{\citenamefont {Akmal}\ and\ \citenamefont
  {Pandharipande}(1997)}]{Akmal:1997ft}%
  \BibitemOpen
  \bibfield  {author} {\bibinfo {author} {\bibfnamefont {A.}~\bibnamefont
  {Akmal}}\ and\ \bibinfo {author} {\bibfnamefont {V.~R.}\ \bibnamefont
  {Pandharipande}},\ }\href {\doibase 10.1103/PhysRevC.56.2261} {\bibfield
  {journal} {\bibinfo  {journal} {Phys. Rev.}\ }\textbf {\bibinfo {volume}
  {C56}},\ \bibinfo {pages} {2261} (\bibinfo {year} {1997})},\ \Eprint
  {http://arxiv.org/abs/nucl-th/9705013} {arXiv:nucl-th/9705013 [nucl-th]}
  \BibitemShut {NoStop}%
\bibitem [{\citenamefont {Dashen}\ \emph {et~al.}(1969)\citenamefont {Dashen},
  \citenamefont {Ma},\ and\ \citenamefont {Bernstein}}]{Dashen:1969}%
  \BibitemOpen
  \bibfield  {author} {\bibinfo {author} {\bibfnamefont {R.}~\bibnamefont
  {Dashen}}, \bibinfo {author} {\bibfnamefont {S.-k.}\ \bibnamefont {Ma}}, \
  and\ \bibinfo {author} {\bibfnamefont {H.~J.}\ \bibnamefont {Bernstein}},\
  }\href {\doibase 10.1103/PhysRev.187.345} {\bibfield  {journal} {\bibinfo
  {journal} {Phys. Rev.}\ }\textbf {\bibinfo {volume} {187}},\ \bibinfo {pages}
  {345} (\bibinfo {year} {1969})}\BibitemShut {NoStop}%
\bibitem [{\citenamefont {Venugopalan}\ and\ \citenamefont
  {Prakash}(1992)}]{Venugopalan:1992hy}%
  \BibitemOpen
  \bibfield  {author} {\bibinfo {author} {\bibfnamefont {R.}~\bibnamefont
  {Venugopalan}}\ and\ \bibinfo {author} {\bibfnamefont {M.}~\bibnamefont
  {Prakash}},\ }\href {\doibase 10.1016/0375-9474(92)90005-5} {\bibfield
  {journal} {\bibinfo  {journal} {Nucl. Phys.}\ }\textbf {\bibinfo {volume}
  {A546}},\ \bibinfo {pages} {718} (\bibinfo {year} {1992})}\BibitemShut
  {NoStop}%
\bibitem [{\citenamefont {Huovinen}\ and\ \citenamefont
  {Petreczky}(2018)}]{Huovinen:2017ogf}%
  \BibitemOpen
  \bibfield  {author} {\bibinfo {author} {\bibfnamefont {P.}~\bibnamefont
  {Huovinen}}\ and\ \bibinfo {author} {\bibfnamefont {P.}~\bibnamefont
  {Petreczky}},\ }\href {\doibase 10.1016/j.physletb.2017.12.001} {\bibfield
  {journal} {\bibinfo  {journal} {Phys. Lett.}\ }\textbf {\bibinfo {volume}
  {B777}},\ \bibinfo {pages} {125} (\bibinfo {year} {2018})},\ \Eprint
  {http://arxiv.org/abs/1708.00879} {arXiv:1708.00879 [hep-ph]} \BibitemShut
  {NoStop}%
\bibitem [{\citenamefont {Horowitz}\ and\ \citenamefont
  {Schwenk}(2006)}]{Horowitz:2005zv}%
  \BibitemOpen
  \bibfield  {author} {\bibinfo {author} {\bibfnamefont {C.~J.}\ \bibnamefont
  {Horowitz}}\ and\ \bibinfo {author} {\bibfnamefont {A.}~\bibnamefont
  {Schwenk}},\ }\href {\doibase 10.1016/j.physletb.2006.05.055} {\bibfield
  {journal} {\bibinfo  {journal} {Phys. Lett.}\ }\textbf {\bibinfo {volume}
  {B638}},\ \bibinfo {pages} {153} (\bibinfo {year} {2006})},\ \Eprint
  {http://arxiv.org/abs/nucl-th/0507064} {arXiv:nucl-th/0507064 [nucl-th]}
  \BibitemShut {NoStop}%
\bibitem [{\citenamefont {Skyrme}(1958)}]{Skyrme:1958}%
  \BibitemOpen
  \bibfield  {author} {\bibinfo {author} {\bibfnamefont {T.}~\bibnamefont
  {Skyrme}},\ }\href {\doibase https://doi.org/10.1016/0029-5582(58)90345-6}
  {\bibfield  {journal} {\bibinfo  {journal} {Nuclear Physics}\ }\textbf
  {\bibinfo {volume} {9}},\ \bibinfo {pages} {615 } (\bibinfo {year}
  {1958})}\BibitemShut {NoStop}%
\bibitem [{\citenamefont {Steiner}\ \emph {et~al.}(2005)\citenamefont
  {Steiner}, \citenamefont {Prakash}, \citenamefont {Lattimer},\ and\
  \citenamefont {Ellis}}]{Steiner:2004fi}%
  \BibitemOpen
  \bibfield  {author} {\bibinfo {author} {\bibfnamefont {A.~W.}\ \bibnamefont
  {Steiner}}, \bibinfo {author} {\bibfnamefont {M.}~\bibnamefont {Prakash}},
  \bibinfo {author} {\bibfnamefont {J.~M.}\ \bibnamefont {Lattimer}}, \ and\
  \bibinfo {author} {\bibfnamefont {P.~J.}\ \bibnamefont {Ellis}},\ }\href
  {\doibase 10.1016/j.physrep.2005.02.004} {\bibfield  {journal} {\bibinfo
  {journal} {Phys. Rept.}\ }\textbf {\bibinfo {volume} {411}},\ \bibinfo
  {pages} {325} (\bibinfo {year} {2005})},\ \Eprint
  {http://arxiv.org/abs/nucl-th/0410066} {arXiv:nucl-th/0410066 [nucl-th]}
  \BibitemShut {NoStop}%
\bibitem [{\citenamefont {Dutra}\ \emph {et~al.}(2012)\citenamefont {Dutra},
  \citenamefont {Lourenco}, \citenamefont {Sa~Martins}, \citenamefont
  {Delfino}, \citenamefont {Stone},\ and\ \citenamefont
  {Stevenson}}]{Dutra:2012mb}%
  \BibitemOpen
  \bibfield  {author} {\bibinfo {author} {\bibfnamefont {M.}~\bibnamefont
  {Dutra}}, \bibinfo {author} {\bibfnamefont {O.}~\bibnamefont {Lourenco}},
  \bibinfo {author} {\bibfnamefont {J.~S.}\ \bibnamefont {Sa~Martins}},
  \bibinfo {author} {\bibfnamefont {A.}~\bibnamefont {Delfino}}, \bibinfo
  {author} {\bibfnamefont {J.~R.}\ \bibnamefont {Stone}}, \ and\ \bibinfo
  {author} {\bibfnamefont {P.~D.}\ \bibnamefont {Stevenson}},\ }\href {\doibase
  10.1103/PhysRevC.85.035201} {\bibfield  {journal} {\bibinfo  {journal} {Phys.
  Rev.}\ }\textbf {\bibinfo {volume} {C85}},\ \bibinfo {pages} {035201}
  (\bibinfo {year} {2012})},\ \Eprint {http://arxiv.org/abs/1202.3902}
  {arXiv:1202.3902 [nucl-th]} \BibitemShut {NoStop}%
\bibitem [{\citenamefont {Tews}\ \emph {et~al.}(2013)\citenamefont {Tews},
  \citenamefont {Kr{\"u}ger}, \citenamefont {Hebeler},\ and\ \citenamefont
  {Schwenk}}]{Tews:2012fj}%
  \BibitemOpen
  \bibfield  {author} {\bibinfo {author} {\bibfnamefont {I.}~\bibnamefont
  {Tews}}, \bibinfo {author} {\bibfnamefont {T.}~\bibnamefont {Kr{\"u}ger}},
  \bibinfo {author} {\bibfnamefont {K.}~\bibnamefont {Hebeler}}, \ and\
  \bibinfo {author} {\bibfnamefont {A.}~\bibnamefont {Schwenk}},\ }\href
  {\doibase 10.1103/PhysRevLett.110.032504} {\bibfield  {journal} {\bibinfo
  {journal} {Phys. Rev. Lett.}\ }\textbf {\bibinfo {volume} {110}},\ \bibinfo
  {pages} {032504} (\bibinfo {year} {2013})},\ \Eprint
  {http://arxiv.org/abs/1206.0025} {arXiv:1206.0025 [nucl-th]} \BibitemShut
  {NoStop}%
\bibitem [{\citenamefont {Gandolfi}\ \emph {et~al.}(2014)\citenamefont
  {Gandolfi}, \citenamefont {Carlson}, \citenamefont {Reddy}, \citenamefont
  {Steiner},\ and\ \citenamefont {Wiringa}}]{Gandolfi:2013baa}%
  \BibitemOpen
  \bibfield  {author} {\bibinfo {author} {\bibfnamefont {S.}~\bibnamefont
  {Gandolfi}}, \bibinfo {author} {\bibfnamefont {J.}~\bibnamefont {Carlson}},
  \bibinfo {author} {\bibfnamefont {S.}~\bibnamefont {Reddy}}, \bibinfo
  {author} {\bibfnamefont {A.~W.}\ \bibnamefont {Steiner}}, \ and\ \bibinfo
  {author} {\bibfnamefont {R.~B.}\ \bibnamefont {Wiringa}},\ }\href {\doibase
  10.1140/epja/i2014-14010-5} {\bibfield  {journal} {\bibinfo  {journal} {Eur.
  Phys. J.}\ }\textbf {\bibinfo {volume} {A50}},\ \bibinfo {pages} {10}
  (\bibinfo {year} {2014})},\ \Eprint {http://arxiv.org/abs/1307.5815}
  {arXiv:1307.5815 [nucl-th]} \BibitemShut {NoStop}%
\bibitem [{\citenamefont {Lo}(2017)}]{Lo:2017}%
  \BibitemOpen
  \bibfield  {author} {\bibinfo {author} {\bibfnamefont {P.~M.}\ \bibnamefont
  {Lo}},\ }\href {\doibase 10.1140/epjc/s10052-017-5106-0} {\bibfield
  {journal} {\bibinfo  {journal} {The European Physical Journal C}\ }\textbf
  {\bibinfo {volume} {77}},\ \bibinfo {pages} {533} (\bibinfo {year}
  {2017})}\BibitemShut {NoStop}%
\bibitem [{\citenamefont {Hoferichter}\ \emph {et~al.}(2016)\citenamefont
  {Hoferichter}, \citenamefont {Ruiz~de Elvira}, \citenamefont {Kubis},\ and\
  \citenamefont {Mei{\ss}ner}}]{Hoferichter:2015hva}%
  \BibitemOpen
  \bibfield  {author} {\bibinfo {author} {\bibfnamefont {M.}~\bibnamefont
  {Hoferichter}}, \bibinfo {author} {\bibfnamefont {J.}~\bibnamefont {Ruiz~de
  Elvira}}, \bibinfo {author} {\bibfnamefont {B.}~\bibnamefont {Kubis}}, \ and\
  \bibinfo {author} {\bibfnamefont {U.-G.}\ \bibnamefont {Mei{\ss}ner}},\
  }\href {\doibase 10.1016/j.physrep.2016.02.002} {\bibfield  {journal}
  {\bibinfo  {journal} {Phys. Rept.}\ }\textbf {\bibinfo {volume} {625}},\
  \bibinfo {pages} {1} (\bibinfo {year} {2016})},\ \Eprint
  {http://arxiv.org/abs/1510.06039} {arXiv:1510.06039 [hep-ph]} \BibitemShut
  {NoStop}%
\bibitem [{\citenamefont {Meissner}\ \emph {et~al.}(2002)\citenamefont
  {Meissner}, \citenamefont {Oller},\ and\ \citenamefont
  {Wirzba}}]{Meissner:2001gz}%
  \BibitemOpen
  \bibfield  {author} {\bibinfo {author} {\bibfnamefont {U.~G.}\ \bibnamefont
  {Meissner}}, \bibinfo {author} {\bibfnamefont {J.~A.}\ \bibnamefont {Oller}},
  \ and\ \bibinfo {author} {\bibfnamefont {A.}~\bibnamefont {Wirzba}},\ }\href
  {\doibase 10.1006/aphy.2002.6244} {\bibfield  {journal} {\bibinfo  {journal}
  {Annals Phys.}\ }\textbf {\bibinfo {volume} {297}},\ \bibinfo {pages} {27}
  (\bibinfo {year} {2002})},\ \Eprint {http://arxiv.org/abs/nucl-th/0109026}
  {arXiv:nucl-th/0109026 [nucl-th]} \BibitemShut {NoStop}%
\bibitem [{\citenamefont {Hoferichter}\ \emph {et~al.}(2015)\citenamefont
  {Hoferichter}, \citenamefont {Ruiz~de Elvira}, \citenamefont {Kubis},\ and\
  \citenamefont {Mei{\ss}ner}}]{Hoferichter:2015tha}%
  \BibitemOpen
  \bibfield  {author} {\bibinfo {author} {\bibfnamefont {M.}~\bibnamefont
  {Hoferichter}}, \bibinfo {author} {\bibfnamefont {J.}~\bibnamefont {Ruiz~de
  Elvira}}, \bibinfo {author} {\bibfnamefont {B.}~\bibnamefont {Kubis}}, \ and\
  \bibinfo {author} {\bibfnamefont {U.-G.}\ \bibnamefont {Mei{\ss}ner}},\
  }\href {\doibase 10.1103/PhysRevLett.115.192301} {\bibfield  {journal}
  {\bibinfo  {journal} {Phys. Rev. Lett.}\ }\textbf {\bibinfo {volume} {115}},\
  \bibinfo {pages} {192301} (\bibinfo {year} {2015})},\ \Eprint
  {http://arxiv.org/abs/1507.07552} {arXiv:1507.07552 [nucl-th]} \BibitemShut
  {NoStop}%
\bibitem [{\citenamefont {{Perego}}\ \emph {et~al.}(2019)\citenamefont
  {{Perego}}, \citenamefont {{Bernuzzi}},\ and\ \citenamefont
  {{Radice}}}]{Perego:2019}%
  \BibitemOpen
  \bibfield  {author} {\bibinfo {author} {\bibfnamefont {A.}~\bibnamefont
  {{Perego}}}, \bibinfo {author} {\bibfnamefont {S.}~\bibnamefont
  {{Bernuzzi}}}, \ and\ \bibinfo {author} {\bibfnamefont {D.}~\bibnamefont
  {{Radice}}},\ }\href {\doibase 10.1140/epja/i2019-12810-7} {\bibfield
  {journal} {\bibinfo  {journal} {European Physical Journal A}\ }\textbf
  {\bibinfo {volume} {55}},\ \bibinfo {eid} {124} (\bibinfo {year} {2019})},\
  \Eprint {http://arxiv.org/abs/1903.07898} {arXiv:1903.07898 [gr-qc]}
  \BibitemShut {NoStop}%
\bibitem [{\citenamefont {Janka}(2012)}]{Janka:2012wk}%
  \BibitemOpen
  \bibfield  {author} {\bibinfo {author} {\bibfnamefont {H.-T.}\ \bibnamefont
  {Janka}},\ }\href {\doibase 10.1146/annurev-nucl-102711-094901} {\bibfield
  {journal} {\bibinfo  {journal} {Ann. Rev. Nucl. Part. Sci.}\ }\textbf
  {\bibinfo {volume} {62}},\ \bibinfo {pages} {407} (\bibinfo {year} {2012})},\
  \Eprint {http://arxiv.org/abs/1206.2503} {arXiv:1206.2503 [astro-ph.SR]}
  \BibitemShut {NoStop}%
\bibitem [{\citenamefont {Burrows}\ and\ \citenamefont
  {Lattimer}(1986)}]{Burrows:1986me}%
  \BibitemOpen
  \bibfield  {author} {\bibinfo {author} {\bibfnamefont {A.}~\bibnamefont
  {Burrows}}\ and\ \bibinfo {author} {\bibfnamefont {J.~M.}\ \bibnamefont
  {Lattimer}},\ }\href {\doibase 10.1086/164405} {\bibfield  {journal}
  {\bibinfo  {journal} {Astrophys. J.}\ }\textbf {\bibinfo {volume} {307}},\
  \bibinfo {pages} {178} (\bibinfo {year} {1986})}\BibitemShut {NoStop}%
\bibitem [{\citenamefont {Pons}\ \emph {et~al.}(1999)\citenamefont {Pons},
  \citenamefont {Reddy}, \citenamefont {Prakash}, \citenamefont {Lattimer},\
  and\ \citenamefont {Miralles}}]{Pons:1998mm}%
  \BibitemOpen
  \bibfield  {author} {\bibinfo {author} {\bibfnamefont {J.~A.}\ \bibnamefont
  {Pons}}, \bibinfo {author} {\bibfnamefont {S.}~\bibnamefont {Reddy}},
  \bibinfo {author} {\bibfnamefont {M.}~\bibnamefont {Prakash}}, \bibinfo
  {author} {\bibfnamefont {J.~M.}\ \bibnamefont {Lattimer}}, \ and\ \bibinfo
  {author} {\bibfnamefont {J.~A.}\ \bibnamefont {Miralles}},\ }\href {\doibase
  10.1086/306889} {\bibfield  {journal} {\bibinfo  {journal} {Astrophys. J.}\
  }\textbf {\bibinfo {volume} {513}},\ \bibinfo {pages} {780} (\bibinfo {year}
  {1999})},\ \Eprint {http://arxiv.org/abs/astro-ph/9807040}
  {arXiv:astro-ph/9807040 [astro-ph]} \BibitemShut {NoStop}%
\bibitem [{\citenamefont {Roberts}\ and\ \citenamefont
  {Reddy}(2017{\natexlab{a}})}]{Roberts:2017pns}%
  \BibitemOpen
  \bibfield  {author} {\bibinfo {author} {\bibfnamefont {L.~F.}\ \bibnamefont
  {Roberts}}\ and\ \bibinfo {author} {\bibfnamefont {S.}~\bibnamefont
  {Reddy}},\ }\enquote {\bibinfo {title} {Neutrino signatures from young
  neutron stars},}\ in\ \href {\doibase 10.1007/978-3-319-21846-5_5} {\emph
  {\bibinfo {booktitle} {Handbook of Supernovae}}},\ \bibinfo {editor} {edited
  by\ \bibinfo {editor} {\bibfnamefont {A.~W.}\ \bibnamefont {Alsabti}}\ and\
  \bibinfo {editor} {\bibfnamefont {P.}~\bibnamefont {Murdin}}}\ (\bibinfo
  {publisher} {Springer International Publishing},\ \bibinfo {address} {Cham},\
  \bibinfo {year} {2017})\ pp.\ \bibinfo {pages} {1605--1635}\BibitemShut
  {NoStop}%
\bibitem [{\citenamefont {Sekiguchi}\ \emph {et~al.}(2011)\citenamefont
  {Sekiguchi}, \citenamefont {Kiuchi}, \citenamefont {Kyutoku},\ and\
  \citenamefont {Shibata}}]{Sekiguchi:2011zd}%
  \BibitemOpen
  \bibfield  {author} {\bibinfo {author} {\bibfnamefont {Y.}~\bibnamefont
  {Sekiguchi}}, \bibinfo {author} {\bibfnamefont {K.}~\bibnamefont {Kiuchi}},
  \bibinfo {author} {\bibfnamefont {K.}~\bibnamefont {Kyutoku}}, \ and\
  \bibinfo {author} {\bibfnamefont {M.}~\bibnamefont {Shibata}},\ }\href
  {\doibase 10.1103/PhysRevLett.107.051102} {\bibfield  {journal} {\bibinfo
  {journal} {Phys. Rev. Lett.}\ }\textbf {\bibinfo {volume} {107}},\ \bibinfo
  {pages} {051102} (\bibinfo {year} {2011})},\ \Eprint
  {http://arxiv.org/abs/1105.2125} {arXiv:1105.2125 [gr-qc]} \BibitemShut
  {NoStop}%
\bibitem [{\citenamefont {Gasser}\ and\ \citenamefont
  {Zarnauskas}(2010)}]{Gasser:2010}%
  \BibitemOpen
  \bibfield  {author} {\bibinfo {author} {\bibfnamefont {J.}~\bibnamefont
  {Gasser}}\ and\ \bibinfo {author} {\bibfnamefont {G.}~\bibnamefont
  {Zarnauskas}},\ }\href {\doibase
  https://doi.org/10.1016/j.physletb.2010.08.021} {\bibfield  {journal}
  {\bibinfo  {journal} {Physics Letters B}\ }\textbf {\bibinfo {volume}
  {693}},\ \bibinfo {pages} {122 } (\bibinfo {year} {2010})}\BibitemShut
  {NoStop}%
\bibitem [{\citenamefont {Mahan}(2000)}]{Mahan:2000}%
  \BibitemOpen
  \bibfield  {author} {\bibinfo {author} {\bibfnamefont {G.~D.}\ \bibnamefont
  {Mahan}},\ }\href@noop {} {\emph {\bibinfo {title} {Many Particle Physics,
  Third Edition}}}\ (\bibinfo  {publisher} {Plenum},\ \bibinfo {address} {New
  York},\ \bibinfo {year} {2000})\BibitemShut {NoStop}%
\bibitem [{\citenamefont {Horowitz}\ and\ \citenamefont
  {Wehrberger}(1991)}]{Horowitz:1990it}%
  \BibitemOpen
  \bibfield  {author} {\bibinfo {author} {\bibfnamefont {C.~J.}\ \bibnamefont
  {Horowitz}}\ and\ \bibinfo {author} {\bibfnamefont {K.}~\bibnamefont
  {Wehrberger}},\ }\href {\doibase 10.1016/0375-9474(91)90745-R} {\bibfield
  {journal} {\bibinfo  {journal} {Nucl. Phys.}\ }\textbf {\bibinfo {volume}
  {A531}},\ \bibinfo {pages} {665} (\bibinfo {year} {1991})}\BibitemShut
  {NoStop}%
\bibitem [{\citenamefont {Reddy}\ \emph {et~al.}(1999)\citenamefont {Reddy},
  \citenamefont {Prakash}, \citenamefont {Lattimer},\ and\ \citenamefont
  {Pons}}]{Reddy:1998hb}%
  \BibitemOpen
  \bibfield  {author} {\bibinfo {author} {\bibfnamefont {S.}~\bibnamefont
  {Reddy}}, \bibinfo {author} {\bibfnamefont {M.}~\bibnamefont {Prakash}},
  \bibinfo {author} {\bibfnamefont {J.~M.}\ \bibnamefont {Lattimer}}, \ and\
  \bibinfo {author} {\bibfnamefont {J.~A.}\ \bibnamefont {Pons}},\ }\href
  {\doibase 10.1103/PhysRevC.59.2888} {\bibfield  {journal} {\bibinfo
  {journal} {Phys. Rev.}\ }\textbf {\bibinfo {volume} {C59}},\ \bibinfo {pages}
  {2888} (\bibinfo {year} {1999})},\ \Eprint
  {http://arxiv.org/abs/astro-ph/9811294} {arXiv:astro-ph/9811294 [astro-ph]}
  \BibitemShut {NoStop}%
\bibitem [{\citenamefont {Roberts}\ and\ \citenamefont
  {Reddy}(2017{\natexlab{b}})}]{Roberts:2017}%
  \BibitemOpen
  \bibfield  {author} {\bibinfo {author} {\bibfnamefont {L.~F.}\ \bibnamefont
  {Roberts}}\ and\ \bibinfo {author} {\bibfnamefont {S.}~\bibnamefont
  {Reddy}},\ }\href {\doibase 10.1103/PhysRevC.95.045807} {\bibfield  {journal}
  {\bibinfo  {journal} {Phys. Rev. C}\ }\textbf {\bibinfo {volume} {95}},\
  \bibinfo {pages} {045807} (\bibinfo {year} {2017}{\natexlab{b}})}\BibitemShut
  {NoStop}%
\bibitem [{\citenamefont {{Cutler}}\ \emph {et~al.}(1990)\citenamefont
  {{Cutler}}, \citenamefont {{Lindblom}},\ and\ \citenamefont
  {{Splinter}}}]{Cutler:1990}%
  \BibitemOpen
  \bibfield  {author} {\bibinfo {author} {\bibfnamefont {C.}~\bibnamefont
  {{Cutler}}}, \bibinfo {author} {\bibfnamefont {L.}~\bibnamefont
  {{Lindblom}}}, \ and\ \bibinfo {author} {\bibfnamefont {R.~J.}\ \bibnamefont
  {{Splinter}}},\ }\href {\doibase 10.1086/169370} {\bibfield  {journal}
  {\bibinfo  {journal} {\apj}\ }\textbf {\bibinfo {volume} {363}},\ \bibinfo
  {pages} {603} (\bibinfo {year} {1990})}\BibitemShut {NoStop}%
\bibitem [{\citenamefont {Alford}\ \emph {et~al.}(2018)\citenamefont {Alford},
  \citenamefont {Bovard}, \citenamefont {Hanauske}, \citenamefont {Rezzolla},\
  and\ \citenamefont {Schwenzer}}]{Alford:2017rxf}%
  \BibitemOpen
  \bibfield  {author} {\bibinfo {author} {\bibfnamefont {M.~G.}\ \bibnamefont
  {Alford}}, \bibinfo {author} {\bibfnamefont {L.}~\bibnamefont {Bovard}},
  \bibinfo {author} {\bibfnamefont {M.}~\bibnamefont {Hanauske}}, \bibinfo
  {author} {\bibfnamefont {L.}~\bibnamefont {Rezzolla}}, \ and\ \bibinfo
  {author} {\bibfnamefont {K.}~\bibnamefont {Schwenzer}},\ }\href {\doibase
  10.1103/PhysRevLett.120.041101} {\bibfield  {journal} {\bibinfo  {journal}
  {Phys. Rev. Lett.}\ }\textbf {\bibinfo {volume} {120}},\ \bibinfo {pages}
  {041101} (\bibinfo {year} {2018})},\ \Eprint
  {http://arxiv.org/abs/1707.09475} {arXiv:1707.09475 [gr-qc]} \BibitemShut
  {NoStop}%
\bibitem [{\citenamefont {Sawyer}(1989)}]{Sawyer:1989}%
  \BibitemOpen
  \bibfield  {author} {\bibinfo {author} {\bibfnamefont {R.~F.}\ \bibnamefont
  {Sawyer}},\ }\href {\doibase 10.1103/PhysRevD.39.3804} {\bibfield  {journal}
  {\bibinfo  {journal} {Phys. Rev. D}\ }\textbf {\bibinfo {volume} {39}},\
  \bibinfo {pages} {3804} (\bibinfo {year} {1989})}\BibitemShut {NoStop}%
\bibitem [{\citenamefont {Alford}\ and\ \citenamefont
  {Harris}(2019)}]{Alford:2019qtm}%
  \BibitemOpen
  \bibfield  {author} {\bibinfo {author} {\bibfnamefont {M.~G.}\ \bibnamefont
  {Alford}}\ and\ \bibinfo {author} {\bibfnamefont {S.~P.}\ \bibnamefont
  {Harris}},\ }\href {\doibase 10.1103/PhysRevC.100.035803} {\bibfield
  {journal} {\bibinfo  {journal} {Phys. Rev.}\ }\textbf {\bibinfo {volume}
  {C100}},\ \bibinfo {pages} {035803} (\bibinfo {year} {2019})},\ \Eprint
  {http://arxiv.org/abs/1907.03795} {arXiv:1907.03795 [nucl-th]} \BibitemShut
  {NoStop}%
\bibitem [{\citenamefont {Alford}\ \emph
  {et~al.}(2019{\natexlab{a}})\citenamefont {Alford}, \citenamefont
  {Harutyunyan},\ and\ \citenamefont {Sedrakian}}]{Alford:2019kdw}%
  \BibitemOpen
  \bibfield  {author} {\bibinfo {author} {\bibfnamefont {M.}~\bibnamefont
  {Alford}}, \bibinfo {author} {\bibfnamefont {A.}~\bibnamefont {Harutyunyan}},
  \ and\ \bibinfo {author} {\bibfnamefont {A.}~\bibnamefont {Sedrakian}},\
  }\href@noop {} {\  (\bibinfo {year} {2019}{\natexlab{a}})},\ \Eprint
  {http://arxiv.org/abs/1907.04192} {arXiv:1907.04192 [astro-ph.HE]}
  \BibitemShut {NoStop}%
\bibitem [{\citenamefont {Alford}\ \emph
  {et~al.}(2019{\natexlab{b}})\citenamefont {Alford}, \citenamefont {Fore},
  \citenamefont {Harris},\ and\ \citenamefont {Reddy}}]{pionbulkviscoisity}%
  \BibitemOpen
  \bibfield  {author} {\bibinfo {author} {\bibfnamefont {M.~G.}\ \bibnamefont
  {Alford}}, \bibinfo {author} {\bibfnamefont {B.}~\bibnamefont {Fore}},
  \bibinfo {author} {\bibfnamefont {S.~P.}\ \bibnamefont {Harris}}, \ and\
  \bibinfo {author} {\bibfnamefont {S.}~\bibnamefont {Reddy}},\ }\href@noop {}
  {\  (\bibinfo {year} {2019}{\natexlab{b}})}\BibitemShut {NoStop}%
\bibitem [{\citenamefont {Wellenhofer}\ \emph {et~al.}(2015)\citenamefont
  {Wellenhofer}, \citenamefont {Holt},\ and\ \citenamefont
  {Kaiser}}]{Wellenhofer:2015qba}%
  \BibitemOpen
  \bibfield  {author} {\bibinfo {author} {\bibfnamefont {C.}~\bibnamefont
  {Wellenhofer}}, \bibinfo {author} {\bibfnamefont {J.~W.}\ \bibnamefont
  {Holt}}, \ and\ \bibinfo {author} {\bibfnamefont {N.}~\bibnamefont
  {Kaiser}},\ }\href {\doibase 10.1103/PhysRevC.92.015801} {\bibfield
  {journal} {\bibinfo  {journal} {Phys. Rev.}\ }\textbf {\bibinfo {volume}
  {C92}},\ \bibinfo {pages} {015801} (\bibinfo {year} {2015})},\ \Eprint
  {http://arxiv.org/abs/1504.00177} {arXiv:1504.00177 [nucl-th]} \BibitemShut
  {NoStop}%
\end{thebibliography}%

\end{document}